\documentclass{emulateapj}
\usepackage{natbib}
\usepackage{amssymb}
\usepackage{setspace}
\usepackage{graphicx}
\usepackage{rotating}
\usepackage{color}
\submitted{submitted to the Astrophysical Journal, Under Review}

\def\CIVdblt{{\rm C~}\kern 0.1em{\sc iv}~$\lambda\lambda 1548, 1550$}
\def\MgIIdblt{{\rm Mg~}\kern 0.1em{\sc ii}~$\lambda\lambda 2796, 2803$}
\def\NVdblt{{\rm N}\kern 0.1em{\sc v}~$\lambda\lambda 1238, 1242$}  
\def\OVIdblt{{\rm O}\kern 0.1em{\sc vi}~$\lambda\lambda 1032, 1038$}
\def\SiIVdblt{{\rm Si~}\kern 0.1em{\sc iv}~$\lambda\lambda1394, 1403$}
\def\AlIIIdblt{{\rm Al~}\kern 0.1em{\sc iii}~$\lambda\lambda1855,1863$}
\def\FeIIdblt{{\rm Fe~}\kern 0.1em{\sc ii}~$\lambda\lambda 2383, 2600$}
\def\NeVIIIdblt{{\rm Ne~}\kern 0.1em{\sc viii}~$\lambda\lambda 770, 780$}

\def\NeVIII{\hbox{{\rm Ne~}\kern 0.1em{\sc viii}}}
\def\OI{\hbox{{\rm O~}\kern 0.1em{\sc i}}}
\def\OII{\hbox{{\rm O~}\kern 0.1em{\sc ii}}}
\def\OIII{\hbox{{\rm O~}\kern 0.1em{\sc iii}}}
\def\OIV{\hbox{{\rm O~}\kern 0.1em{\sc iv}}}
\def\OV{\hbox{{\rm O~}\kern 0.1em{\sc v}}}
\def\OVI{\hbox{{\rm O~}\kern 0.1em{\sc vi}}}
\def\OVII{\hbox{{\rm O~}\kern 0.1em{\sc vii}}}
\def\OVIII{\hbox{{\rm O~}\kern 0.1em{\sc viii}}}
\def\NIII{\hbox{{\rm N~}\kern 0.1em{\sc iii}}}
\def\NIV{\hbox{{\rm N~}\kern 0.1em{\sc iv}}}
\def\NVII{\hbox{{\rm N~}\kern 0.1em{\sc vii}}}
\def\CIII{\hbox{{\rm C~}\kern 0.1em{\sc iii}}}
\def\SiIII{\hbox{{\rm Si~}\kern 0.1em{\sc iii}}}
\def\SVI{\hbox{{\rm S~}\kern 0.1em{\sc vi}}}
\def\NeIX{\hbox{{\rm Ne~}\kern 0.1em{\sc ix}}}

\def\AlII{\hbox{{\rm Al~}\kern 0.1em{\sc ii}}}
\def\AlIII{\hbox{{\rm Al~}\kern 0.1em{\sc iii}}}
\def\CaI{\hbox{{\rm Ca}\kern 0.1em{\sc i}}}
\def\CaII{\hbox{{\rm Ca}\kern 0.1em{\sc ii}}}
\def\CrII{\hbox{{\rm Cr}\kern 0.1em{\sc ii}}}
\def\CII{\hbox{{\rm C~}\kern 0.1em{\sc ii}}}
\def\CIII{\hbox{{\rm C~}\kern 0.1em{\sc iii}}}
\def\CIV{\hbox{{\rm C~}\kern 0.1em{\sc iv}}}
\def\CV{\hbox{{\rm C}\kern 0.1em{\sc v}}}
\def\H{\hbox{{\rm H}}}
\def\HI{\hbox{{\rm H~}\kern 0.1em{\sc i}}}
\def\HII{\hbox{{\rm H~}\kern 0.1em{\sc ii}}}
\def\Lya{\hbox{{\rm Ly}\kern 0.1em$\alpha$}}
\def\Lyb{\hbox{{\rm Ly}\kern 0.1em$\beta$}}
\def\Lyg{\hbox{{\rm Ly}\kern 0.1em$\gamma$}}
\def\Lyth{\hbox{{\rm Ly}\kern 0.1em$\theta$}}
\def\Lyfive{\hbox{{\rm Ly}\kern 0.1em$5$}}
\def\Lysix{\hbox{{\rm Ly}\kern 0.1em$6$}}
\def\Lyseven{\hbox{{\rm Ly}\kern 0.1em$7$}}
\def\Lyeight{\hbox{{\rm Ly}\kern 0.1em$8$}}
\def\Lynine{\hbox{{\rm Ly}\kern 0.1em$9$}}
\def\Lyten{\hbox{{\rm Ly}\kern 0.1em$10$}}
\def\HeI{\hbox{{\rm He}\kern 0.1em{\sc i}}}
\def\HeII{\hbox{{\rm He}\kern 0.1em{\sc ii}}}
\def\FeI{\hbox{{\rm Fe~}\kern 0.1em{\sc i}}}
\def\FeII{\hbox{{\rm Fe~}\kern 0.1em{\sc ii}}}
\def\FeIII{\hbox{{\rm Fe~}\kern 0.1em{\sc iii}}}
\def\MnII{\hbox{{\rm Mn}\kern 0.1em{\sc ii}}}
\def\MgI{\hbox{{\rm Mg~}\kern 0.1em{\sc i}}}
\def\MgII{\hbox{{\rm Mg~}\kern 0.1em{\sc ii}}}
\def\MgIII{\hbox{{\rm Mg~}\kern 0.1em{\sc iii}}}
\def\MgIV{\hbox{{\rm Mg~}\kern 0.1em{\sc iv}}}
\def\NaI{\hbox{{\rm Na}\kern 0.1em{\sc i}}}
\def\NV{\hbox{{\rm N}\kern 0.1em{\sc v}}}
\def\NII{\hbox{{\rm N}\kern 0.1em{\sc ii}}}
\def\NIII{\hbox{{\rm N}\kern 0.1em{\sc iii}}}
\def\OVI{\hbox{{\rm O}\kern 0.1em{\sc vi}}}
\def\SiII{\hbox{{\rm Si~}\kern 0.1em{\sc ii}}}
\def\SiIII{\hbox{{\rm Si~}\kern 0.1em{\sc iii}}}
\def\SiIV{\hbox{{\rm Si~}\kern 0.1em{\sc iv}}}
\def\SII{\hbox{{\rm S}\kern 0.1em{\sc ii}}}
\def\SIII{\hbox{{\rm S}\kern 0.1em{\sc iii}}}
\def\SIV{\hbox{{\rm S}\kern 0.1em{\sc iv}}}
\def\TiII{\hbox{{\rm Ti}\kern 0.1em{\sc ii}}}
\def\ZnII{\hbox{{\rm Zn}\kern 0.1em{\sc ii}}}
\newcommand{\kms}{\hbox{km~s$^{-1}$}}
\newcommand{\cmsq}{\hbox{cm$^{-2}$}}
\newcommand{\cc}{\hbox{cm$^{-3}$}}
\def\kms{\hbox{km~s$^{-1}$}}      
\def\cmsq{\hbox{cm$^{-2}$}}
\def\cc{\hbox{cm$^{-3}$}}
\def\etal{et~al.\ }

\begin{document}

\title{Cosmic Origins Spectrograph Detection of {\NeVIII} \\ Tracing Warm - Hot Gas Towards PKS~$0405-123$\altaffilmark{1}}

\author{Anand Narayanan, Blair D. Savage, Bart P. Wakker\altaffilmark{2}, Charles W. Danforth,  Yangsen Yao, \\ Brian A. Keeney, J. Michael Shull\altaffilmark{3}, Kenneth R. Sembach\altaffilmark{4}, \\ Cynthia S. Froning, \& James C. Green\altaffilmark{3}}

\altaffiltext{1}{Based on observations with the NASA/ESA {\it Hubble Space Telescope}, obtained at the Space Telescope Science Institute, which is operated by the Association of Universities for Research in Astronomy, Inc., under NASA contract NAS 05-26555, and the NASA-CNES/ESA {\it Far Ultraviolet SpectroscopicExplorer} mission, operated by the Johns Hopkins University, supported by NASA contract NAS 05-32985.}
\altaffiltext{2}{Department of Astronomy, The University of Wisconsin-Madison, 5534 Sterling Hall, 475 N. Charter Street, Madison WI 53706-1582, USA, Email: anand, wakker, wakker@astro.wisc.edu}
\altaffiltext{3}{CASA, Department of Astrophysical and Planetary Sciences, University of Colorado, 389-UCB, Boulder, CO 80309, USA.}
\altaffiltext{4}{Space Telescope Science Institute, 3700 San Martin Drive, Baltimore, MD 21218, USA.}
%\subjectheadings{galaxies: halos, intergalactic medium, quasars: absorption lines, quasars: individual: PKS 0405-123, ultraviolet: general}

\begin{abstract}

We report on the detection of {\NeVIII} in the $HST$/Cosmic Origins Spectrograph spectrum of the intervening absorption system at $z = 0.495096$ towards PKS~$0405-123$ ($z_{em} = 0.5726$). The high $S/N$ COS spectrum also covers absorption from {\HI}, {\CIII}, {\OIII}, {\OIV} and {\OVI} associated with this multiphase system. The {\NeVIII} is detected with high significance in both lines of the doublet, with integrated column densities of log~$N_a(\NeVIII~770) = 13.96~{\pm}~0.06$ and log~$N_a(\NeVIII~780) = 14.08~{\pm}~0.07$. We find the origin of {\NeVIII} consistent with collisionally ionized gas at $T \sim 5 \times 10^5$~K with a large baryonic column density of $N(\H) \sim 10^{19} - 10^{20}$~{\cmsq}. The  metallicity in the {\NeVIII} gas phase is estimated to be [Ne/H]~$\sim -0.6~{\pm}~0.3$~dex. The intermediate ions such as {\CIII}, {\OIII}, {\OIV} and {\HI} are consistent with photoionization in lower ionization gas at $T \sim 10^4$~K. The {\OV} and {\OVI} in this absorber can have contributions from both the photoionized and collisionally ionized gas phases. The absorber is at $|\Delta v| = 180$~{\kms} systematic velocity and $\rho = 110~h^{-1}_{70}$~kpc projected separation from a $M_R = -19.6$ galaxy of extended morphology. The collisionally ionized gas at $T \sim 5 \times 10^5$~K detected in {\NeVIII} and {\OVI} points to an origin in multiphase gas embedded in the {\it hot} halo of the galaxy, or in a nearby WHIM structure. The high sensitivity UV spectroscopy afforded by COS has opened up new opportunities for discovering large reservoirs of {\it missing baryons} in the low-$z$ universe through the detection of {\NeVIII} systems.

\end{abstract}

\section{Introduction}

Absorption-line spectroscopy of distant quasars have yielded a complete census of the properties and distribution of baryons at the early epochs of the universe. At $z \gtrsim 3$, almost all of the baryons reside in the space in-between galaxies and galaxy clusters. This intergalactic medium (IGM) at high-$z$ is almost entirely in the form of gas photoionized and heated to $T \sim 10^4$~K \citep[e.g.][]{fukugita98}. The photoionized IGM manifests in the spectra of quasars as the {\Lya} forest. From those early epochs, the process of galaxy formation as well as the continued expansion of the universe is expected to have drastically altered the phase composition of much of the baryons \citep{cen99, dave01, cen06}. Observational efforts to detect these baryons in the present universe have so far not succeeded in identifying their most dominant reservoirs \citep{bregman07}. 

In the $z \sim 0$ universe, collapsed objects such as galaxies, groups and clusters account for only $\lesssim 10$\% of the cosmic baryon budget (Fukugita \& Peebles 2004). More than $90$\% of the baryons are still outside of galaxies. However, unlike high-$z$, the fraction of intergalactic baryons in the photoionized phase far from galaxies has significantly declined. Observationally, this is evident from the decline in the redshift distribution of {\Lya} absorbers \citep{kim97,weymann98,penton00}. Estimates suggest that in the present universe, the {\Lya} forest contribution to the baryon density ($\Omega_{\Lya}$) is only $\gtrsim 30$\% (Penton {\etal}2000, 2004; Lehner {\etal}2007, Danforth \& Shull 2008). This implies that a larger fraction of the intergalactic baryons must exist in a separate gas phase, not included in galaxies and also not dominantly photoionized.  Numerical simulations of structure formation predict that most of these baryons are in highly ionized gas structures at temperatures in the range $T \sim 10^5 - 10^7$~K and densities of $n_{\H} \sim (0.1 - 10) \times 10^{-5}$~{\cc} \citep{cen99, dave01, valageas02, cen06}. Frequently referred to as the {\it Warm Hot Intergalactic Medium} (WHIM), this gas phase was an outcome of collisional ionization through heating in gravitational shocks when intergalactic matter fell into the potential wells of collapsed dark matter as structures grew hierarchically \citep[e.g.][]{cen99}. Detecting the WHIM and characterizing its physical properties remains one of the most important themes in observational cosmology. For extensive reviews on the search for WHIM in the low-$z$ universe see \citet{bregman07} \&  \citet{prochaska09}.

UV absorption-line spectroscopy is presently the most promising approach for the detection and characterization of the WHIM. The diffuse nature predicted for the WHIM gas makes detection via emission scarcely possible. Also, attempts at observing the $T \gtrsim 10^6$~K WHIM in X-ray absorption at $z > 0$ have not been successful largely due to the insufficient sensitivity and resolution of the current generation of instruments. In the UV, the principal absorption-line tracers of collisionally ionized gas are {\OVIdblt}, {\NeVIIIdblt} and broad {\Lya} (BLA). In gas in collisional ionization equilibrium, {\OVI} reaches its peak ionization fraction at  $T \sim 3 \times 10^5$~K. The strong {\OVIdblt} doublet lines can thus be a sensitive probe of gas at those collisionally ionized temperatures \citep{tripp00, savage02, danforth08, narayanan10a, narayanan10b}. However, {\OVI} is also produced in low density environments under the influence of a strong radiation field through pure photoionization at $T \sim 10^4$~K \citep{savage02, prochaska04, lehner06, thom08b, tripp08, oppenheimer09}. In the case of most {\OVI} absorption systems, the complexity seen in absorption prevent a reliable assessment of the process that dominates the ionization. 

In comparison, the {\NeVIII}~$\lambda\lambda~770.409, 780.324$ doublet lines \citep[$f_{770} = 0.103, f_{780} = 0.0505$;][]{verner94} are better tracers of collisionally ionized gas. Even though the cosmic abundance of neon is less compared to oxygen \citep[(Ne/O)$_\odot = 0.17$;][]{asplund09}, detectable amounts of {\NeVIII} can be present in {\OVI} absorbing gas at $T \sim (0.5 - 1.5) \times 10^6$~K. The temperature range corresponds to the {\it warm} phase of the WHIM. In collisional ionization equilibrium (CIE), the {\NeVIII} ionization fraction peaks at $T = 7 \times 10^5$~K. The first clear detection ($> 3~\sigma$) of {\NeVIII} doublet lines for an intervening absorber at low-$z$ was reported by Savage {\etal}(2005), in the high-$S/N$ $FUSE$ spectrum of the metal line system at $z = 0.20701$ towards HE~$0226-4110$. The column density of {\NeVIII} and its ratio with {\OVI} were consistent with an origin in collisionally ionized gas at $T = 5.4 \times 10^5$~K with a substantial baryonic column of $N(\H) \sim 8 \times 10^{19}$~{\cmsq}. The intermediate ions (such as {\CIII}, {\OIII}, {\OIV}, {\SiIII}) and strong {\HI} absorption in this system were created in a separate gas phase at $T \sim 2 \times 10^4$~K through photoionization. The second detection of {\NeVIII} was reported by \citet{narayanan09} in an intervening absorption-line system at $z = 0.32566$ in the $FUSE$ spectrum of the quasar 3C 263. The redshifted wavelengths of {\OVIdblt} lines in this absorber fell outside the wavelength coverage of $FUSE$. Nonetheless, the detection of ions such as {\OIII}, {\OIV} and {\NV} in the $FUSE$ spectrum pointed to the presence of both photoionized and collisionally ionized regions in the absorber, traced by the intermediate ions and {\NeVIII} respectively. The detection of {\NeVIII} required the presence of gas with $T \gtrsim 10^5$~K, corresponding to the temperature anticipated for the {\it warm} phase of the WHIM. 

Here we report on another instance of {\NeVIII} detection in an intervening absorption system in the low-$z$ universe. The {\NeVIIIdblt} doublet lines are detected at high significance in the high $S/N$ $HST$/Cosmic Origins Spectrograph (COS) spectrum of the UV bright quasar PKS~$0405-123$. The detection distinctly points to the presence of {\it warm} collisionally ionized gas in a structure with a very large column density [$N(\H) \gtrsim 10^{19}$~{\cmsq}]  of baryons. This is third in a series of COS detections of collisionally ionized gas in the regions surrounding galaxies, the other two being \citet{savage10} and \citet{narayanan10b}.

\section{Overview of Previous Work}

The PKS~$0405-123$ sight line was previously observed by the $FUSE$ and $HST$/STIS instruments at spectral resolutions of $\sim 20$~{\kms} and $\sim 7$~{\kms}. Lower resolution FOS observations (FWHM $\sim 230$~{\kms}) also exist for this target. A number of authors have published results on the detection and analysis of hydrogen and metal-line absorbers along this sight line \citep{bahcall93, jannuzi98, prochaska04, williger06, lehner07, tripp08, thom08a, thom08b, howk09}. The most complete analysis of the ionization and chemical abundances in the $z = 0.495096$ absorber is given in Howk {\etal}(2009). A thorough investigation of the physical conditions was made possible by the combined $FUSE$ and STIS coverage of several important low and high ionization metal species and hydrogen for this absorber. 

In the $FUSE$ spectrum, the {\NeVIIIdblt} lines were non-detections with a 3~$\sigma$ upper limit on the equivalent width and column density of $W_r(\NeVIII~770) < 29$~m{\AA} and log~$N_a(\NeVIII) < 13.73$ in the rest-frame of the absorber (Howk {\etal}2009). In the absence of {\NeVIII}, Howk {\etal}(2009) were able to determine a single-phase photoionization model that could simultaneously explain the column densities of all ions, including {\OVI}. The constraints from the photoionization models, particularly the metallicity, were found to be dependent on the nature of the ionizing spectrum. In the best-fit models, an extragalactic ionizing background dictated by QSOs predict carbon and oxygen abundances of $-0.15$~dex, whereas including the contribution of ionizing photons from star forming galaxies, the abundances drop to $-0.62$~dex. The single-phase photoionization models derived by Howk {\etal}(2009) predict densities of $n_{\H} \sim 5 \times 10^{-5}$~{\cc}, baryonic column densities of log~$N(\H) \sim 18.5$, a photoionization equilibrium temperature of $T \sim 28,000$~K and a physical size of $L \sim 20$~kpc for the absorbing region. Howk {\etal}(2009) also do not rule out the possibility of the absorber having multiple gas phases, with at least one phase of predominantly photoionized gas ($T \sim 10^4$~K) and another wamer phase ($T \sim 10^5$~K) that is collisionally ionized. The absence of {\NeVIII} in the $FUSE$ spectrum was accepted as evidence for the lack of $T \sim (0.5 - 3) \times 10^6$~K {\it warm - hot} gas associated with the absorber. The detection of {\NeVIII} in the higher sensitivity COS spectrum is a crucial new result, as it convincingly demonstrates that we are observing a substantial baryonic column of collisionally ionized gas, possibly the WHIM or the ionized halo of a galaxy. 
 
\section{COS Observations}

The COS spectrum for PKS~$0405-123$ presented here is a combination of observations from the HST Early Release Program of August 2009 (Program ID: 11508) and GTO observations by the science team from December 2009 (Program ID: 11541). The details of the separate COS integrations are listed in Table 1. The exposures were retrieved from the HST archive and reduced in a uniform fashion using the most current CalCOS pipeline software (ver 2.11). The reduced data were flux calibrated. The design capabilities of $HST$/COS are described in detail by \citet{green01, froning09} and in the updated COS Instrument Handbook \citep{cos2010}. The inflight performance of COS is discussed by \citet{osterman10} and in the numerous instrument science reports on the STScI COS website\footnote{http://www.stsci.edu/hst/cos/documents/isrs}. \citet{savage10} have used these COS observations of PKS~$0405-123$ to study the properties of the highly ionized plasma in the Lyman Limit system at $z = 0.1671$.

The separate G130M and G160M grating integrations were combined together in flux units weighted by their respective exposure time using the custom coaddition routine developed by Charles Danforth and the COS GTO team\footnote{http://casa.colorado.edu/$\sim$danforth/science/cos/costools.html}. A full description of this routine is given in \citep{danforth10b}. In brief, the routine cross-correlates and corrects the different exposures for velocity misalignments, before combining the fluxes. Since the FUV COS spectra are not flat-fielded, a procedure is built into the routine to {\it psuedo-flat field} each exposure to remove narrow fixed pattern features at the $> 10$\% depth (in flux units), most of which are repetitive grid-wire shadows. Applying this automated procedure does not result in the complete removal of every fixed pattern feature. Nonetheless, tests on several COS spectra have shown that the procedure is successful in removing periodic instrumental artifacts, thereby improving the overall sensitivity of the spectrum. The final coadded spectrum has wavelength coverage from  1135~{\AA} to 1796~{\AA. The $S/N$ ratios (per 18~{\kms} resolution element) of this combined final spectrum at 1200~{\AA}, 1400~{\AA}, and 1600~{\AA} are 20, 15, and 18 respectively. We also correct the COS spectrum for wavelength zero-point offsets. Weak low-ionization Milky Way ISM lines observed in the G130M and G160M integrations were used to check the reliability of the COS wavelength calibration. Based on the {\HI} emission data given in \citet{lockman95} we estimate the heliocentric velocity of the ISM absorption in the direction of PKS~$0405-123$ to be $v_\mathrm{HELIO} = 19.9$~{\kms}. To make the ISM lines in the PKS~$0405-123$ COS spectrum have average heliocentric velocities of 20~{\kms}, we applied velocity offsets of $+6$~{\kms} to the G130M and $+22$~{\kms} to the G160M data. Residual errors in the wavelength calibration correspond to $\sim 5$~{\kms}.

Ghavamian {\etal}(2009) has determined the resolving power of the spectrograph by detailed modeling of the line spread function (LSF) at various wavelengths. The spectral resolution is found to be wavelength dependent with values in the range $\lambda/\Delta\lambda \sim 16,000 - 21,000$ for the G130M and G160M gratings, where $\Delta\lambda$ refers to the width at half strength of the LSF which has broad wings containing $\sim 20$\% of the LSF area. The resolution is maximum at near-UV wavelengths and declines monotonically towards lower wavelengths. 

\section{Properties of the {\NeVIII} System}

In this section we describe the observed properties of the $z = 0.495096$ absorption system. In Figure 1, we display the continuum normalized absorption profiles in the rest-frame of the absorber and in Table 1, we list the basic line measurements. For consistency, we adopt the same redshift for the system as given in Howk {\etal}(2009), which was based on the centroid of the {\CIII}~$\lambda 977$ profile from STIS. The COS spectrum for this system shows absorption from {\HI}, {\OVIdblt}, {\OIV}~$\lambda 788$, {\CIII}~$\lambda 977$, {\OIII}~$\lambda 833$, and {\NeVIIIdblt}. In addition, it also covers wavelengths where absorption from {\CII}, {\NII}, {\SiII} and {\FeIII} are expected. For the line measurements, we use the apparent optical depth (AOD) method of Savage \& Sembach (1991). For {\Lyb}, {\OVI}, {\OIV} and {\CIII} we also apply Voigt profile models to estimate the column density, Doppler width and velocity associated with individual components. The fit models are shown in Figure 2. The profiles were fit with Voigt functions using the fitting routine of Fitzpatrick \& Spitzer (1997). In this process, the model profiles were convolved with the COS instrumental spread function at the redshifted wavelength of the line.  It is important that the specific line spread function be used to fit the profile in order to minimize the impact of the LSF non-Gaussian wings on observed line profiles \citep[see][for a detailed discussion]{coslsf}. For completeness, we also show in Figure 1, the strong {\OV}~$\lambda 630$ profile recorded at intermediate resolution by $FUSE$ and the {\Lya} as seen by the low-resolution FOS. This is the first direct detection of {\OV} in the IGM at low-$z$ \citep{prochaska04, howk09}, and its absorption profile agrees well with the component structure for {\OVI} seen in the COS data. 

The {\OVIdblt} lines in the high $S/N$ COS spectrum agree generally with the higher resolution, lower $S/N$ observation from STIS. The {\OVI} absorption is spread over $\Delta v \sim 250$~{\kms} and shows sub-component structure. In the STIS spectrum, a comparison of the apparent column density $N_a(v)$ profiles for the 1032~{\AA} and 1038~{\AA} lines had shown extra absorption in the 1038~{\AA} line at $v \sim 30$~{\kms}, which was labeled as an unidentified interloping absorber (Howk {\etal}2009). In the COS spectrum, we do not find any marked difference between the velocity components of the {\OVI} doublet lines (see Figure 3). The contaminating feature in {\OVI}~1038~{\AA}, found in the STIS spectrum, is more likely a spectral artifact. 

The $N_a(v)$ comparison of Figure 3 shows that the {\OVI} lines are subjected to saturation that is unresolved by COS. The difference in the integrated apparent column densities of  log~$N_a (\OVI~1032) = 14.39~{\pm}~0.01$, and log~$N_a (\OVI~1038) = 14.48~{\pm}~0.01$ is significant and suggestive of this unresolved saturation. The {\OVI}~1038 column density measurement is within ${\pm}~1~\sigma$ of the same by Howk {\etal}(2009) in the STIS spectrum, and is likely to be a closer representation of the true column density. We correct for the line saturation effect using the  procedure described in Savage \& Sembach (1991), and obtain log~$N_a(\OVI) = 14.48~+~0.09 = 14.57$. The $0.09$~dex saturation correction corresponds to the difference between the logarithmic column densities of the weaker (less saturated) and stronger (more saturated) lines of the {\OVI} doublet. The integrated {\OVI} column density which we adopt in all subsequent analysis is the saturation corrected value of log~$N_a(\OVI) = 14.57~{\pm}~0.05$~dex. 

The multiphase nature of the absorber becomes evident while comparing the apparent column density profiles across the absorption seen in the intermediate ions ({\CIII}, {\OIV}) and the high ions ({\OV}, {\OVI}). The {\OIV} and {\CIII} velocity structures resemble each other, as revealed by the arrangement of their $N_a(v)$ profiles displayed in Figure 3. The absorption from these ions over the $-100$~{\kms}~$\leq v \leq 30$~{\kms} interval has a two component structure centered at velocities of $v \sim 0$~{\kms} and $v \sim -30$~{\kms}. However, the component at $v \sim -30$~{\kms} is not readily seen in {\OVI}. It is likely that the slight asymmetry in the negative velocity portion of the {\OVI} absorption is suggestive of an additional component, perhaps thermally broader than what is seen in {\OIV} and {\CIII}. Additional evidence for the multiphase nature of the absorber is evident from the difference between the weak absorption seen in {\CIII} and {\OIV} compared with the strong absorption in {\OVI} over the velocity interval $30$~{\kms}~$\leq v \leq 80$~{\kms}. This indicates that the gas in this velocity range has higher ionization, compared to the gas at negative velocity, although they are likely kinematically linked to each other. The integrated column densities for {\OIV} and {\CIII} when compared to the STIS measurements suggest unresolved saturation for both lines. Howk {\etal} quote a lower limit of log~$N_a(\OIV) \gtrsim 14.37$ and log~$N(\CIII) = 13.39~{\pm}~0.05$ adopted from component fitting. We adopt the same measurements for these two ions in our ionization analysis. 

In the COS spectrum, {\OIII}~$\lambda 833$ is detected at $> 3$~$\sigma$ significance. The line is very weak and therefore was undetected at the sensitivity afforded by $FUSE$. The velocity range over which {\OIII} shows absorption concurs well with the velocity range of absorption from the higher ionization stages of oxygen. The rest-frame equivalent width $W_r(\OIII~833) = 33~{\pm}~2$~m{\AA} and column density log~$N_a(\OIII) = 13.73~{\pm}~0.03$ that we measure from the COS data are consistent with the upper-limits of $W_r < 36$~m{\AA} and log~$N_a < 13.73$ obtained from $FUSE$ by Howk {\etal}(2009). The errors on the {\OIII} equivalent width and column density do not include an estimated 5~m{\AA} and 0.06~dex systematic error produced by fixed pattern noise. With the detection of {\OIII}, we have column density measurements for four successive ionization stages of oxygen ({\OIII} - {\OVI}), which sets useful constraints on the physical conditions in the absorber. 

The information on {\HI} associated with this absorber was poor in the existing STIS and FOS data. The FOS spectrum showed {\Lya} absorption at the redshift of the system, although the low resolution provided little information on the {\HI} kinematics or the possibility of saturation. The {\Lyb} observed at higher resolution by $STIS$ was poorly detected due to the low $S/N \sim 6$~pixel$^{-1}$ of the recorded spectrum. The COS data offers a factor of 3 higher $S/N$ coverage of {\Lyb}, along with tighter constrains on the {\HI} column density from the  higher order lines of the Lyman series.  The {\Lyb} feature in COS is a clear but weak detection, with a central optical depth of $\tau_\mathrm{max} (v) \sim 0.3$. The {\HI} is kinematically broad, spread over a velocity of $\Delta v \sim 350$~{\kms}. Based on simultaneous fits to the {\Lya} (FOS), {\Lyb} (STIS) and {\Lyg} (STIS, $< 3\sigma$ significance) features, Howk {\etal}(2009) derive a {\HI} column density of log~$N(\HI) = 14.29~{\pm}~0.10$. By integrating the $N_a(v)$ profile of {\Lyb} in the COS spectrum, we estimate the logarithmic total {\HI} column density as $14.21~{\pm}~0.02$, consistent with the STIS measurement.  At the COS resolution, the {\Lyb} appears to have two components, a broader principal component at $v \sim -3$~{\kms} blended with a $v \sim 90$~{\kms} component which could be narrower than the instrumental width. This positive velocity component is weaker and may not contribute significantly to the total {\HI} column density. The profile fit to COS {\Lyb} yields $b(\HI) = 53~{\pm}~1$~{\kms} for the principal $v \sim -3$~{\kms} component. If the broadening of this component is dominantly thermal, then the implied temperature is $T = 1.7 \times 10^5$~K. This component appears to be symmetric with respect to its centroid, although the COS instrumental resolution is not adequate to rule out of blending of closely separated components. 

\subsection{The {\NeVIIIdblt} Detection}

The wavelengths of the redshifted {\NeVIIIdblt} lines are 1151.8~{\AA} and 1166.2~{\AA} respectively. The {\NeVIII}~$\lambda 770$ line is therefore covered only by G130M grating observations at central wavelength settings of $1291$~{\AA} and $1300$~{\AA}. From the various COS integrations for this sight line, five exposures (with $t_{exp} \geq 1000$~sec) have central wavelength of $\lambda_c = 1291$~{\AA}, and six shorter exposures ($t_{exp} \sim 650$~sec) have $\lambda_c = 1300$~{\AA}. Similarly, the {\NeVIII}~$\lambda 780$ line is covered by G130M grating settings of central wavelengths $1291$~{\AA}, $1300$~{\AA}, $1309$~{\AA} and $1318$~{\AA}. 

The \NeVIII} in this system is a relatively weak feature. To prevent gross systematic uncertainties from affecting the validity of the {\NeVIII} detection, we customized the combining of spectra in the wavelength range where absorption from the redshifted {\NeVIIIdblt} lines were expected. We carefully chose for coaddition only those individual exposures for which we could clearly rule out grid wire shadows or other detector artifacts (at the redshifted wavelengths of the {\NeVIII} doublet lines). Choosing different central wavelength grating settings allows the dispersed light to shift across the detector in the dispersion direction. For each setting, this enables the recording of the same wavelength by different regions of detector. By then aligning in detector space the different integrations, it becomes possible to readily identify if the wavelengths corresponding to the {\NeVIIIdblt} lines in each exposure are affected by any fixed pattern structure. In the case of PKS~$0405-123$, the individual integrations had adequate $S/N$ to carry out this selection process with some certainty. 

The spectrum at the wavelength of the {\NeVIII}~$\lambda 770$ line is coaddition of four G130M integrations at the $\lambda_c = 1291$~{\AA} setting, resulting in a total exposure time of 5.8 ksec at $\sim 1151.8$~{\AA} . The IDs of these four exposures are labeled in Table 1. At the {\NeVIII}~$\lambda 780$~{\AA} redshifted wavelength, we found all five G130M integrations with $\lambda_c = 1291$~{\AA} setting free of strong fixed pattern noise features and thus suitable for coaddition. These exposure IDs are also labeled in Table 1. The {\NeVIIIdblt} lines displayed in the system plot of Figure 1 and the line measurements given in Table 2 are based on this revised coaddition. 

By integrating over a 200~{\kms} velocity interval, we measure a rest-frame equivalent width of $W_r = 45~{\pm}~6$~m{\AA} for the {\NeVIII}~$\lambda 770$ line. The significance of the detection is $7.5~\sigma$. The 1~$\sigma$ uncertainty listed here includes both statistical errors and continuum placement error. A more conservative estimate of the detection significance should take into account systematic uncertainty from fixed pattern features that contribute to the noise at the $\sim 5$~m{\AA} level.  Including this systematic error, the  {\NeVIII}~$\lambda 770$ is detected with a significance of 6.6$\sigma$. 

At the location of the redshifted {\NeVIII}~$\lambda 780$ line, we also se absorption that is consistent with the 770~{\AA} line. We measure a rest-equivalent width of $W_r(\NeVIII~780) = 29~{\pm}~5$~m{\AA} for this feature implying a detection significance of $5.8\sigma$. The ratio \linebreak $W_r(\NeVIII~770) / W_r(\NeVIII~780) \sim 1.7$ is close to the expected 2:1 line strength ratio for a doublet, further validating the {\NeVIII} detection. The integrated apparent column densities obtained for the two {\NeVIII} lines are also within 1~$\sigma$ of each other. In subsequent analysis, we use log~$N_a = 13.96~{\pm}~0.06$ measured for the {\NeVIII}~$\lambda 770$ line as the {\NeVIII} column density in this absorber. 

The rest frame equivalent width we obtain for the {\NeVIII}~$\lambda 770$ of  $W_r = 45~{\pm}~6$~m{\AA} is somewhat larger than the $3~\sigma$ upper limit of  $W_r < 29$~m{\AA} obtained  from the lower $S/N$  $FUSE$ observations of  \citet{howk09}. We independently re-measured the $FUSE$ observations integrating over ${\pm}~100$~{\kms} and obtain $W_r = 38~{\pm}~12$~m{\AA},  which implies a  $3~\sigma$ detection.   Our COS measurement is therefore in agreement with our new measurement of the $FUSE$ spectrum. The validity of the COS detection of the {\NeVIII}~$\lambda 770$ line is further supported by the matching line profile observed for the much stronger {\OVI}~$\lambda 1032$ line (see Figure 1a).

\section{Ionization In the Absorber}

Our primary goal is to understand the ionization mechanism responsible for the production of {\NeVIII}. Additionally, we investigate the extent to which other high ions such as {\OV} and {\OVI} are produced in the gas phase traced by {\NeVIII}. We start the ionization discussion by considering the possibility of {\NeVIII} arising in a purely photoionized gas. 

\subsection{Is the {\NeVIII} Produced by Photoionization ?}

We use the photoionization code Cloudy [ver.C08.00, \citet{ferland98}] to solve for time equilibrium models that reproduce $N(\NeVIII)$. We treat the other ionic column densities as upper limits to account for the possibility that the absorber could have a mix of gas phases at different ionization levels. The source of ionization is assumed to be dominated by the extragalactic background radiation at the redshift of the absorber, whose shape and intensity is as modeled by \citep{haardt01}. The UV background which we use has contributions from quasars and young star forming galaxies. The photoionization models are calculated for different ionization parameters\footnote{Ionization parameter is defined as the ratio of the number density of photons with E $ \geq 13.6$~eV to the total hydrogen density} log~$U$ and metallicities, to select the model that best-fits the observations. We assume that the entire {\HI} column density of 14.21~dex which we measure for this system is associated with the photoionized gas that we model. This assumption need not be valid for a multi-phase absorber. Therefore we comment on how lower $N(\HI)$ values in the {\NeVIII} gas would change the ionization predictions. In the models, we have assumed that the relative elemental abundance ratios are solar, with abundances of [C/H]$_\odot$ = -3.57~dex, [O/H]$_\odot$ = -3.31~dex, [Ne/H]$_\odot$ = -4.07~dex as given by \citet{asplund09}.

In Figure 4, we display photoionization predicted ionic column densities for solar metallicity, log~$N(\HI) = 14.21$ and different log~$U$. Recovering the observed {\NeVIII} column density through photoionization requires very high ionization parameter values. The observed log~$[N(\NeVIII)/N(\OVI)] \sim -0.6$ is achieved at log~$U \sim -0.2$ (for solar abundance pattern), which corresponds to a number density of $n_{\H} \sim 8 \times 10^{-6}$~{\cc}, and a total hydrogen column density $N(\H) \sim 6 \times 10^{19}$~{\cmsq}. Such diffuse gas with large total hydrogen column density results in a large physical size of $L \sim 2.5$~Mpc for the photoionized region. If the absorption is from an unvirialized structure, then the large path length would result in line broadening due to Hubble expansion of $v = H(z)~L =  215$~{\kms}, much larger than the velocity width of the absorption seen in {\NeVIII} and {\OVI}. If  the observed {\OVI} column also has contribution from the lower ionization gas phase traced by {\CIII}, {\OIII} and {\OIV}, then log~$[N(\NeVIII)/N(\OVI)] > -0.6$ implying log~$U > -0.2$ and even lower densities leading to larger sizes for the absorber. The constraint on the ionization parameter is strongly dependent on the column density ratio between {\NeVIII} to {\OVI} and does not vary significantly with metallicity or the {\HI} column density. The unrealistic predictions of the models suggest that photoionization cannot be responsible for the production of {\NeVIII} in this absorber. The absorber can still have a photoionized phase where much of the {\HI} and intermediate ion absorptions occur, as shown in Howk {\etal}(2009). Such multiphase nature can explain the clear differences in the component structure between the {\HI}, {\CIII}, and {\OIV} profiles with {\OV} and {\OVI} (see Figure 3).

If the absorber is residing in the halo of a galaxy, the local radiation field created by the galaxy could influence the ionization in the absorber \citep{fox05}. The ionization fractions of {\HI}, and low ions such as {\CII}, {\SiII}, {\OIII} and {\OIV} can be influenced by the flux of UV photons escaping from a star forming galaxy. On the other hand, for the ionization fraction of {\OVI} and {\NeVIII} to be altered, photons with energy $E \geq 113.9$~eV and $E \geq 207.2$~eV are required. The radiative intensities of O and B stars at such high energies is very low due to the significant opacity from He$^+$ ionization edge at $E \geq 54.4$~eV. Thus, the  {\OVI} and {\NeVIII} ionization levels are unlikely to be significantly altered by any local galactic radiation field. 

\subsection{Evidence for $T \sim 5 \times 10^5$~K Gas}

\subsubsection{Collisional Ionization Equilibrium}

If collisional ionization equilibrium (CIE) applies, the apparent column density ratio of log~$[N_a(\NeVIII)/N_a(\OVI)] = -0.61$ will be true for an equilibrium temperature of $T \sim 4.7 \times 10^5$~K (assuming solar abundances). If there is contribution from the photoionized phase to the total {\OVI} column density, then the constraint would be log~$[N(\NeVIII)/N(\OVI)] > -0.61$, which is true for temperatures greater than $4.7 \times 10^5$~K.  Thus, the presence of {\OVI} provides a useful lower limit on the temperature of the gas producing the {\NeVIII} absorption. In gas at $T \geq 4.7 \times 10^5$~K, only a small fraction of the hydrogen would be in the neutral form ($f_{\HI} \lesssim N(\HI)/N(\H) = 6.83 \times 10^{-7}$, Gnat \& Sternberg 2007). Also, the {\HI} absorption arising in such gas would be thermally broadened  to $b(\HI) \gtrsim 88$~{\kms}. This implies that much of the {\Lyb} absorption does not arise from the {\NeVIII} phase, but from lower ionization gas that is predominantly photoionized. 

To derive the total hydrogen content of the absorber and the metallicity in the collisionally ionized gas, the {\HI} column associated with this higher ionization {\NeVIII} phase has to be determined. However, the thermally broadened absorption from the trace neutral hydrogen associated with the {\NeVIII} gas would fall on top of the stronger absorption from the photoionized gas. It is therefore difficult to separate out this shallow and broad component of {\HI} from the observed {\Lyb} profile. Therefore one has to determine a column density upper limit for the broad {\HI} indirectly. In Figure 5, we show the residual {\Lyb} absorption after dividing the {\Lyb} spectrum with the Voigt profile fit to feature. On top of this residual spectrum, we superimpose synthesized {\Lyb} profiles with $b(\HI) = 100$~{\kms} and different column density values. It is evident from the figure that the broad {\HI} associated with the {\NeVIII} gas has to be log~$N(\HI) \lesssim 13.5$ to be consistent with the data at the wavelength position of {\Lyb}. We use this adopted upper limit on {\HI} column density to determine the metallicity and the baryonic content in the {\NeVIII} gas. 

In Figure 6, we display \citet{gnat07} CIE models for log~$N(\HI) = 13.5$. At $T = 4.7 \times 10^5$~K, the observed column densities of {\NeVIII} and {\OVI} are well explained if we assume a metallicity of [Z/H] $= -0.35$.  This CIE model would also account for $\lesssim 30$\% of the {\OV} column, but little {\OIV} or {\OIII}, consistent with their origin in a separate (photoionized) phase. The collisional ionization temperature and {\HI} column density in the CIE model are lower and upper limits on the values they could assume. The observed {\NeVIII} column density can be reproduced even when log~$N(\HI) < 13.5$ by increasing the metallicity to values above $-0.35$~dex. The {\HI} ionization fraction at $T = 4.7 \times 10^5$~K implies that the {\NeVIII} gas is tracing a baryonic column of  log~$N(\H) \geq 19.67$. This lower limit on baryonic column density will be smaller for metallicities higher than $-0.35$~dex. It is worth noting that the metallicity inferred for the collisionally ionized gas is consistent with the metallicity range of $-0.62 \leq$~[Z/H]~$\leq -0.15$ estimated by \citet{howk09} for the photoionized gas phase of this absorber. 

Gas that is heated to very high temperatures can undergo rapid radiative cooling at $T \sim 10^5$~K due to enhanced cooling efficiencies \citep{gnat07}. Over time, the ion fractions in  such gas would begin to show departures from CIE due to faster radiative cooling rate compared to the rate at which recombinations occur. However, for the temperatures that we infer for the {\NeVIII} gas, non-equilibrium ionization effects are minimal. At $T = 4.7 \times 10^5$~K, the {\NeVIII} and {\OVI} fractions in a radiatively cooling gas at solar metallicity are $f_{\NeVIII} = 5.26 \times 10^{-2}$ and $f_{\OVI} = 3.64 \times 10^{-2}$ \citep{gnat07}. These values are only marginally different from the CIE values of $f_{\NeVIII} = 4.87 \times 10^{-2}$ and $f_{\OVI} = 3.72 \times 10^{-2}$, and therefore yield similar predictions for the physical conditions.  

\subsubsection{Collisional and Photoionization Hybrid Models}

The CIE analysis described in the previous section did not take into account the incidence of extragalactic ionizing radiation on the absorber. Even when conditions are conducive for collisional processes to dominate the ionization, a realistic modeling of absorption systems should also take into account the influence on ionization by the isotropic radiation field from quasars and star forming galaxies \citep{danforth06, richter06a, tripp08, danforth10a, narayanan10b}. We use Cloudy to compute simple {\it hybrid} models that allow both collisional and photoionization reactions to contribute towards the ionization in the {\NeVIII} gas phase. The temperature is fixed in Cloudy at $T = 4.7 \times 10^5$~K which is the lower limit on the temperature estimated from the CIE model for the {\NeVIII} gas phase. The model that Cloudy would converge on will have the ionization and recombination reactions in equilibrium, even though the heating and cooling rates will be out of balance due to the fixed temperature. 

In Figure 7, we show {\it hybrid} of photoionization and CIE model curves for the {\NeVIII} gas phase with log~$N(\HI) = 13.5$, T $= 4.7 \times 10^5$~K and different log~$U$ values. Comparing the {\it hybrid} model column density predictions with the pure photoionization models displayed in Figure 4, we find that at log~$U \lesssim -1.0$, the ionization fraction of both {\OVI} and {\NeVIII} are higher (and almost a constant) due to the collision of electrons with ions contributing to the ionization at higher densities (smaller log~$U$). The metallicity in this {\it hybrid} phase can have a range of values between $-0.9 \lesssim$~[Z/H]~$\lesssim -0.3$. From the shape of the $N(\OVI)$ curve in Figure 7, it is evident that the upper limit of [Z/H] $\sim -0.3$~dex is based on the assumption that the {\OVI} is exclusively produced in the {\it hybrid} phase. At this metallicity, $N(\NeVIII)$ and $N(\OVI)$ are simultaneously recovered for log~$U < -2.2$, which corresponds to a density of $n_{\H} \gtrsim 8 \times 10^{-4}$~{\cc}, total hydrogen column density of $N(\H) \lesssim 5 \times 10^{19}$~{\cmsq} and line-of-sight thickness of $L \lesssim 20$~kpc. For lower metallicities, log~$U$ has to increase in order to recover the measured $N(\NeVIII)$. The lower limit on the metallicity is the value at which {\NeVIII} is recovered near its peak in ionization fraction and at a log~$U$ that does not result in an excessively large path length for the absorbing region. In the {\it hybrid} model, those conditions are satisfied for [Z/H] $\sim -0.9$ and log~$U \sim -0.5$ which corresponds to $n_{\H} \sim 8 \times 10^{-6}$~{\cc}, $N(\H) \sim 10^{20}$~{\cmsq} and $L \sim 4$~Mpc. Such a {\it hybrid} model would require the {\OVI} to have significant contribution from the separate photoionized phase traced by the low and intermediate ions. To summarize, within the framework of these simple CIE and photoionization {\it hybrid} models, the {\NeVIII} gas phase has a metallicity of [Z/H] $= -0.6~{\pm}0.3$ and log~$U \lesssim -0.5$, which implies $n_{\H} \lesssim 10^{-5}$~{\cc} and $N(\H) \sim 10^{19} - 10^{20}$~{\cmsq}. The {\NeVIII} is produced via collisional ionization, whereas the {\OVI} can have contribution from a separate gas phase that is predominantly photoionized. 

\section{Galaxies In the Vicinity of the Absorber}

Given the evidence for the presence of $T \sim 5 \times 10^5$~K gas, the location of the absorber with reference to galaxies becomes important. Absence of an optical counterpart to the absorber would strengthen the case for the absorption arising in an unvirialized WHIM structure in the IGM. \citet{chen09} have carried out imaging and spectroscopic search for galaxies in the PKS~$0405-123$ field. Their survey has a $> 60$\% completeness for galaxies brighter than R = 22 out to a separation of 2 arc minute from the line of sight, which corresponds to a projected distance of $\sim 734~h^{-1}_{70}$~kpc at $z = 0.4950$, assuming $\Omega_m = 0.28$ and $\Omega_\Lambda = 0.72$ \citep{wright06}. The survey was 100\% complete for galaxies brighter than R = 20. In this survey, a galaxy with rest-frame R magnitude of M$_R - 5$~log~$h_{70} = -19.6$ is identified close to the {\NeVIII} absorber at a projected distance of $\rho = 110~h^{-1}_{70}$~kpc. The galaxy's spectroscopic redshift is $z = 0.4942$, and thus is at a systemic velocity of $\Delta v = -180$~{\kms} with respect to the {\NeVIII} absorber. No other galaxy brighter than $R = 20$ is identified within a comoving distance of 3.7~$h^{-1}_{70}$~Mpc of the absorber. At this redshift, M$_R = -19.6$ corresponds to a galaxy luminosity of $0.08~L_R^*$ \citep{dahlen05}. 

The galaxy displays an extended morphology in the high spatial resolution HST/WFPC2 image of the field \citep[see Figure 10 of][]{chen09}. The galaxy's spectrum is dominated by strong emission lines from [\OII], [\OIII], {\Lyg} and {\Lyb} indicating activity in the underlying stellar population. The proximity of the {\NeVIII} absorber to the galaxy is an important, though not conclusive, indicator of the absorption possibly arising in the extended {\it hot} halo of the galaxy. In the case of the $z = 0.20701$ {\NeVIII} system towards HE~$0226-4110$ reported by \citet{savage05}, three sub-$L^*$ galaxies were identified within $\rho = 200$~kpc and $\Delta v = 300$~{\kms} of the absorber \citep{mulchaey09}. Both these {\NeVIII} detections also show strong {\OVI} absorption ($W_r > 30$~m{\AA}). It is known from a number of absorber-galaxy studies that {\OVI} systems preferentially arise within the $\rho \lesssim 500$~kpc environment of $L*$ galaxies and even closer to $0.1L^*$ galaxies \citep{sembach04, tumlinson05, stocke06, tripp06, wakker09, chen09}. If {\NeVIII} is always associated with strong {\OVI} systems, then they are also likely to be tracing collisionally ionized gas in circumgalactic environments. Investigating this correlation requires a large sample of {\NeVIII} detections. 

The [Z/H] $\sim -0.6$~dex metallicity derived from the {\it hybrid} model for this {\NeVIII} absorber is more consistent with a halo origin rather than in the canonical WHIM structure tracing pristine IGM distant from galaxies, in which case lower metallicities are expected \citep{danforth06}. It is however important to bear in mind that the regions where galaxies interface with the IGM are complex environments influenced by a variety of inflow and outflow processes transferring matter and energy between the two systems. This blurs the distinction between what can be characterized as the halo or the IGM. It is difficult to be conclusive about the astrophysical origin of an absorption system based solely on one dimensional information obtained by probing a pencil-beam through such a complex environment. 

\subsection{Summary}

We have reported on the detection of {\NeVIIIdblt} lines in the intervening absorption system at $z = 0.495096$ in the COS high $S/N$ spectrum of the quasar PKS~$0405-123$. This is the third clear detection of this ion in the low-$z$ IGM. The significant results are summarized as follows :

(1)  Both members of the {\NeVIII} doublet in the $z = 0.495096$ multiphase absorber are detected at high significance (above the $5~\sigma$ level) in the COS spectrum. The {\NeVIII}~$770$ line has a rest-frame equivalent width of $W_r = 45~{\pm}~6$~m{\AA} and an apparent column density of log~$N_a =  13.96~{\pm}~0.06$, and the {\NeVIII}~$\lambda 780$ line has $W_r = 29~{\pm}~5$~m{\AA} and log~$N_a = 14.08~{\pm}~0.07$. The {\NeVIII} absorption extends over the same velocity interval as the strong absorption from {\OVIdblt} lines. 

(2) Also seen in the COS spectrum are {\Lyb}, {\CIII}, {\OIII}, {\OIV} and {\OVI} associated with the absorber. The {\OIII} line was a non-detection at the 3$\sigma$ significance level in the $FUSE$ spectrum for this target. We measure $W_r(\OIII~833) = 33~{\pm}~2$~m{\AA} and log~$N_a(\OIII) = 13.73~{\pm}~0.03$. The {\Lyb} absorption was poorly detected in the low $S/N$ STIS spectrum. The higher sensitivity spectrum for {\Lyb} afforded by COS reveals subcomponent structure in {\HI}, with an integrated apparent {\HI} column density of log~$N_a(\HI) = 14.21~{\pm}~0.02$.

(3) The column density of {\NeVIII} cannot be explained by gas that is purely photoionized by the extragalactic ionizing background radiation field. The detection of {\NeVIII} requires collisionally ionized gas with $T \sim 5 \times 10^5$~K . Gas at this temperature is highly ionized with a trace neutral {\HI} fraction of $f_{\HI} = N(\HI)/N(\H) \sim 7 \times 10^{-7}$.  From the {\Lyb} profile, we estimate the {\HI} column in this {\it warm} collisional phase to be log~$N(\HI) \lesssim 13.5$, which suggests that the {\NeVIII} is tracing a gas structure with a total hydrogen column density of $N(\H) \sim 5 \times 10^{19}$~{\cmsq}. 

(4) {\it Hybrid} models that simultaneously take into account both collisional ionization and photoionization reactions predict a metallicity of [Z/H] $= - 0.6~{\pm}~0.3$~dex in the {\NeVIII} gas phase. The density in this gas phase is constrained to $n_{\H} \lesssim 10^{-5}$~{\cc} and total hydrogen column density to $N(\H) \sim 10^{19} - 10^{20}$~{\cmsq}. Even in the {\it hybrid} models the production of {\NeVIII} is through collisional ionization. 

(5) The {\OV} and {\OVI} absorptions are consistent with a multiphase origin, with contributions from both the $T \sim 10^4$~K photoionized gas traced by the {\HI}, {\CIII}, {\OIII} and {\OIV} and the $T \sim 5 \times 10^5$~K collisionally ionized gas phase where the {\NeVIII} is produced. The metallicity for the collisionally ionized gas phase is consistent with the $-0.62 \leq$~[Z/H]~$\leq -0.15$ metallicity range estimated by \citet{howk09} for the photoionized gas phase in this absorber. 

(6) The Chen \& Mulchaey (2009) imaging survey of the PKS~$0405-123$ field has identified a galaxy with rest-frame R magnitude of $M_R - 5$~log~$h_{70} = - 19.6$ ($L \sim 0.08~L_R^*$) at a projected separation of $\rho = 110~h^{-1}_{70}$ kpc and at a systemic velocity of $\Delta v = - 180$~{\kms} from the {\NeVIII} absorber. 
 
 (7) The proximity of the absorber to the galaxy and their modest velocity displacement from each other suggests the possibility of the {\NeVIII} absorption arising in multiphase gas embedded in the {\it hot} halo of the galaxy. The physical properties and metallicity of the absorber are also consistent with an origin in a nearby intergalactic WHIM structure. 
 
 \noindent {\bf Acknowledgments :}~The authors thank the STS-125 team for completing a highly successful {\it Hubble Space Telescope} servicing mission in 2009. We are grateful to Gary Ferland and collaborators for developing the Cloudy photoionization code. We thank Orly Gnat for making the computational data on radiatively cooling models public. This research is supported by the NASA {\it Cosmic Origins Spectrograph} program through a sub-contract to the University of Wisconsin-Madison from the University of Colorado, Boulder. B.P.W acknowledges support from NASA grant NNX-07AH426. This research has made use of the NASA/IPAC Extragalactic Database (NED) which is operated by the Jet Propulsion Laboratory, California Institute of Technology, under contract with the National Aeronautics and Space Administration.

\pagebreak

%%%%%%%%%%%%%%%%%%%%%%%%%%%%%%%%%%%%%%%%%%%%%%%%%%%%%
%%%%%%%%%%%%%%    TABLES    %%%%%%%%%%%%%%%%%%%%%%%%%%%%%%

\clearpage
%\begin{landscape}
\begin{deluxetable}{lcccccc}
\tabletypesize{\scriptsize} 
\tablewidth{0pt}
\tablecaption{\textsc{COS Observations of PKS~$0405-123$}}
\tablehead{
\colhead{HST ID} &
\colhead{Date of Observation} &
\colhead{Grating} &
\colhead{FP-POS} &
\colhead{Central Wavelength} &
\colhead{Wavelength Range} &
\colhead{Exposure Duration} \\
\colhead{ } &
\colhead{(yyyy:mm:dd)} &
\colhead{ } &
\colhead{ } &
\colhead{(\AA)} &
\colhead{(\AA)} &
\colhead{(sec)} 
}
\startdata
LACB51010	&	2009-08-31		&	G130M	&	3	&	1291	&	$1136 - 1430$	&	984$^{a,b}$\\
LACB51020	&	2009-08-31		&	G130M	&	3	&	1300	&	$1147 - 1440$	&	650	\\
LACB51030	&	2009-08-31		&	G130M	&	3	&	1309	&	$1156 - 1449$	&	1908	\\
LACB51040	&	2009-08-31		&	G130M	&	3	&	1300	&	$1147 - 1440$	&	650	\\
LACB51050	&	2009-08-31		&	G130M	&	3	&	1318	&	$1165 - 1459$	&	1908	\\
LACB51060	&	2009-08-31		&	G130M	&	3	&	1300	&	$1147 - 1440$	&	650	\\
LACB51070	&	2009-08-31		&	G130M	&	3	&	1327	& 	$1175 - 1468$	&	1908	\\
LB6822010	&	2009-12-21		&	G160M	& 	3	&	1589	& 	$1401 - 1761$	&	2167 \\
LB6822020	&	2009-12-21		&	G160M	&	3	&	1600	& 	$1412 - 1772$	&	2965 \\
LB6822030	&	2009-12-21		&	G160M	&	3	&	1611	& 	$1424 - 1784$	&	2965 \\
LB6822040	&	2009-12-21		&	G160M	&	3	&	1623	& 	$1436 - 1796$	&	2965 \\
LB6823010	&	2009-12-21		&	G130M	& 	1	&	1291	&	$1141 - 1435$	&	1000$^{b}$ \\
		&	2009-12-21		&	G130M	&	2	&	1291	&	$1138 - 1432$	&	1400$^{a,b}$\\
		&	2009-12-21		&	G130M	&	3	&	1291	&	$1136 - 1430$	&	1000$^{a,b}$ \\
		&	2009-12-21		&	G130M	&	4	&	1291	&	$1133 - 1427$	&	1430$^{a,b}$
\enddata
\tablecomments{ Column (1) is the HST ID for the respective data set, column (2) shows the date of observation, column (3) lists the choice of grating, column (4) gives the FP position used with the grating setting, column (5) shows the grating central wavelength setting, column (6) gives the wavelength range covered under each setting and column (7) lists the duration of integration. 
\\
\\
\noindent HST IDs dated 2009/08/31 are $HST$ Early Release Observations (ERO) with program ID: 11508, and those dated 2009/12/21 are GTO observations with program ID: 11541 (P.I. James Green). 
\\
\\
\noindent $^a$ The data sets used in the coaddition to produce the spectrum at the redshifted wavelength of {\NeVIII}~$\lambda 770$.
\\
\\
\noindent $^b$ The data sets used in the coaddition to produce the spectrum at the redshifted wavelength of {\NeVIII}~$\lambda 780$.} 
\label{tab:tab1}
\end{deluxetable}
%\end{landscape}

\begin{deluxetable}{lcccccr}
\tablecaption{\textsc{PKS0405-123 $z = 0.495096$ {\OVI} - {\NeVIII} Absorber}}
\tablehead{
\colhead{Line} &
\colhead{$W_r$} &
\colhead{$v$} &
\colhead{log~$[N~(\cmsq)]$} &
\colhead{$b$} &
\colhead{$[-v, +v]$}  & 
\colhead{Method} \\ 
\colhead{ } &
\colhead{(m\AA)} &
\colhead{(\kms)} &
\colhead{dex} &
\colhead{(\kms)} &
\colhead{(\kms)} &
\colhead{ }
}
\startdata
{\Lya}			&  $493~{\pm}~40$&	$8~{\pm}~5$	&	$14.03~{\pm}~0.05$	&	$161~{\pm}~13$	&	[-240, 320]		&	AOD	\\
\\
{\Lyb}		& 	$108~{\pm}~6$ 	& 	$4~{\pm}~2$ 	& 	$14.21~{\pm}~0.02$ 	& 	$85~{\pm}~5$ 	& 	[-150, 150] 		& 	AOD \\
\\
{\Lyb} 		&  		...		& 	 $-3~{\pm}~3$ 	& 	$14.09~{\pm}~0.03$ 	& 	$53~{\pm}~1$ 	&  				& 	Fit 	\\
{ } 		&  		...		& 	$90~{\pm}~4$ 	& 	$13.44~{\pm}~0.09$ 	& 	$21~{\pm}~2$ 	&  				& 	 	\\
\\
{\Lyg}		& 	$52~{\pm}~4$ 	& 	$14~{\pm}~8$ 	& 	$14.36~{\pm}~0.04$ 	& 	$93~{\pm}~12$ 	& 	[-150, 150] 		& 	AOD 	\\
\\
{\OVI}~$1032$ 	& $223~{\pm}~4$ 	& 	$-1~{\pm}~1$ 	& 	$14.39~{\pm}~0.01$ 	& 	$54~{\pm}~4$ 	& 	[-100, 150] 		& 	AOD 	\\
\\
{\OVI}~$1032$ 	& 	...  		&  	$-6~{\pm}~2$ 	& 	$14.29~{\pm}~0.02$ 	& 	$35~{\pm}~1$ 	&  				& 	Fit 	\\
{ } 			&  	...  		&  	$45~{\pm}~2$ 	& 	$13.80~{\pm}~0.07$ 	& 	$19~{\pm}~2$ 	&   				& 	         \\
\\
{\OVI}~$1038$	&  $153~{\pm}~5$	&	$-2~{\pm}~4$	&	$14.48~{\pm}~0.01$	&	$50~{\pm}~4$	&	[-70, 105]		&	AOD	\\
\\
{\NeVIII}~$770$	&  $45~{\pm}~6$	&	$-8~{\pm}~5$	&	$13.96~{\pm}~0.06$	&	$70~{\pm}~4$	&	[-100, 100]		&	AOD  \\
\\
{\NeVIII}~$780$	&  $29~{\pm}~5$	&	$-13~{\pm}~11$	&	$14.08~{\pm}~0.07$	&	$78~{\pm}~7$	&	[-100, 100]		&	AOD \\
\\
{\OIV}~$788$	& $108~{\pm}~4$	&	$-9~{\pm}~2$	&	$14.35~{\pm}~0.01$	&	$52~{\pm}~7$	& 	[-100, 150]		&	AOD \\
\\
{\OIV}~$788$	&	...		&	$-32~{\pm}~3$	&	$13.73~{\pm}~0.11$	&	$8~{\pm}~2$	&				& 	Fit 	\\
{ }			&	...		&	$-3~{\pm}~2$	&	$14.09~{\pm}~0.06$	&	$13~{\pm}~1$	&				&	    	\\
{ }			&	...		&	$16~{\pm}~8$	&	$13.97~{\pm}~0.08$	&	$62~{\pm}~1$	&				&		\\
\\
{\OV}~$630$	&  $188~{\pm}~18$&	$12~{\pm}~4$	&	$ > 14.3$			&	$54~{\pm}~8$	&	[-100, 150]		&	AOD	\\
\\
{\CIII}~$977$	&  $76~{\pm}~5$	&	$-5~{\pm}~2$	&	$13.13~{\pm}~0.03$	&	$55~{\pm}~7$	&	[-100, 150]			&	AOD	\\
\\
{\CIII}~$977$	&	...		&	$-32~{\pm}~3$	&	$12.52~{\pm}~0.07$	&	$12~{\pm}~2$	&				&	Fit 	\\
{ }			&	...		&	$4~{\pm}~1$	&	$12.94~{\pm}~0.03$	&	$11~{\pm}~1$	&				&		\\
{ }			&	...		&	$63~{\pm}~3$	&	$12.40~{\pm}~0.07$	&	$20~{\pm}~1$	&				&		\\
\\
{\OIII}~$833$	&  $33~{\pm}~2$	&	$-26~{\pm}~6$	&	$13.73~{\pm}~0.03$	&	$53~{\pm}~3$	&	[-100, 55]		&	3$\sigma$ \\
\\
{\CII}~$1036$	&  $< 16$		&		...		&	$< 13.2$			&		...		&	[-100, 150]		&	3$\sigma$ \\
\\
{\NIII}~$990$	&  $< 13$		&		...		&	$< 13.1$			&		...		&	[-100, 150]		&	3$\sigma$ \\ 
\\
{\NII}~$1084$	&  $< 54$		&		...		&	$< 12.9$			&		...		&	[-100, 150]		&	3$\sigma$ \\
\\
{\OIII}~$833$	&  $< 49$		&		...		&	$< 13.9$			&		...		&	[-100, 150]		&	3$\sigma$ \\
\\
{\SiII}~$1190$	&   $< 42$		&		...		&	$< 13.1$			&		...		&	[-100, 150]		&	3$\sigma$ \\
\\
{\FeIII}~$1123$	&  $< 25$		&		...		&	$< 13.5$			&		...		&	[-100, 150]		&	3$\sigma$
\enddata
\\
\\
\tablecomments{The {\Lya} is FOS data. The {\OV}~$\lambda~630$~{\AA} is FUSE data. The {\Lyg} could be a 3$\sigma$ upper limit. 
\\
\\
The different $N$ values for {\OVI} from the 1032, and 1038 suggests significant saturation for both lines. To first approximation, the true $N$ value is $N(\OVI) = N(\OVI~1038) + \Delta N$. Thus, $N(\OVI) = 14.48 + 0.09 = 14.57$. The error would be larger, $\sim 0.05$~dex. 
\\
\\
The {\NeVIII}~$770$~{\AA} line is a coaddition of four exposures (1 post-focus 1291 exposure and 3 GTO 1291 exposures). The {\NeVIII}~$780$~{\AA} line is a coaddition of five exposures (1 post-focus 1291 exposure and 4 GTO 1291 exposures).}
\label{tab:tab2}
\end{deluxetable}

%%%%%%%%%%%%%%%%%%%%%%%%%%%%%%%%%%%%%%%%%%%%%%%%%%%%%
%%%%%%%%%%%%%%    FIGURES    %%%%%%%%%%%%%%%%%%%%%%%%%%%%%%

\clearpage
\begin{figure*}
\begin{center}
\includegraphics[scale=0.7]{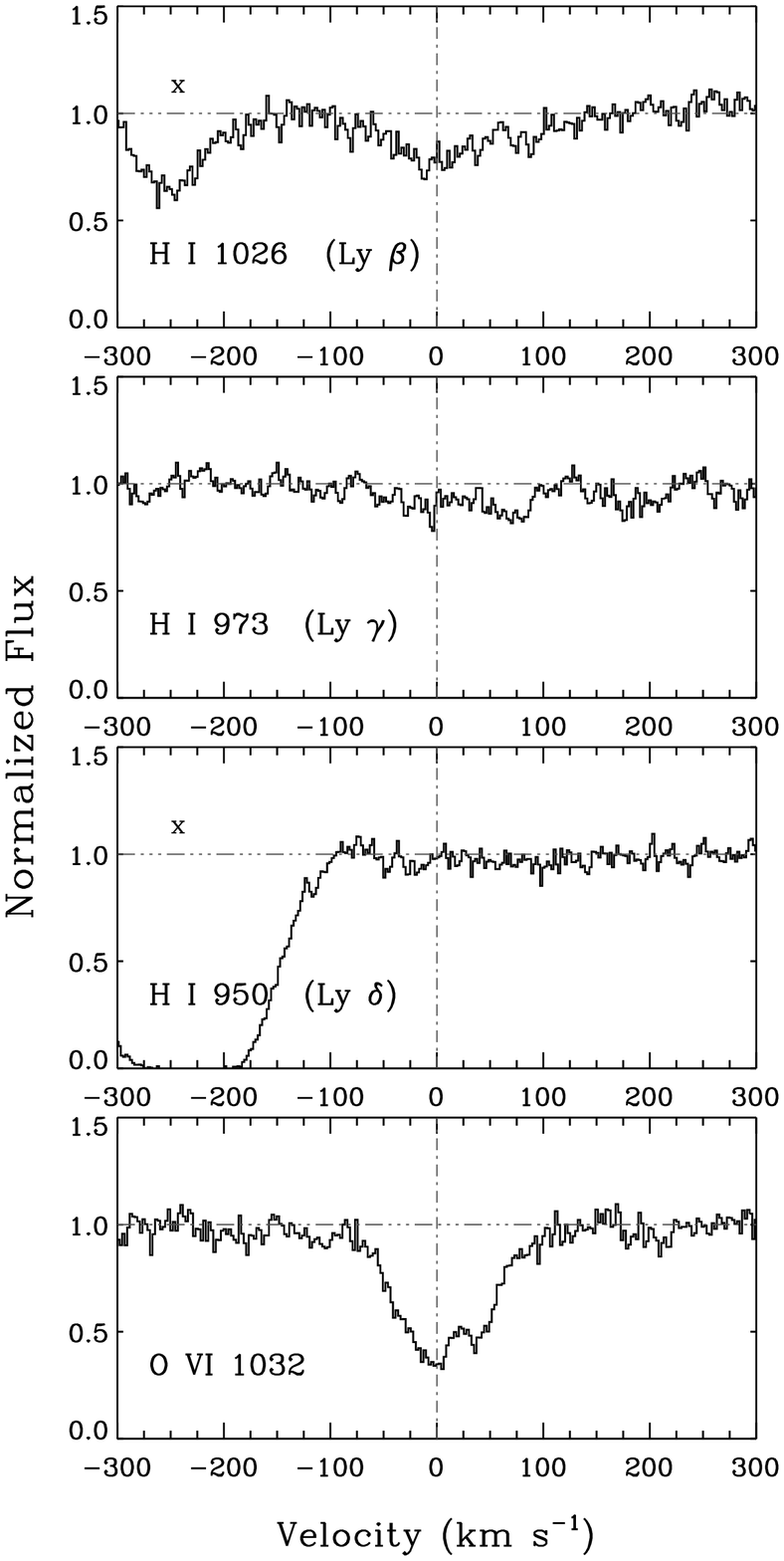}
\includegraphics[scale=0.7]{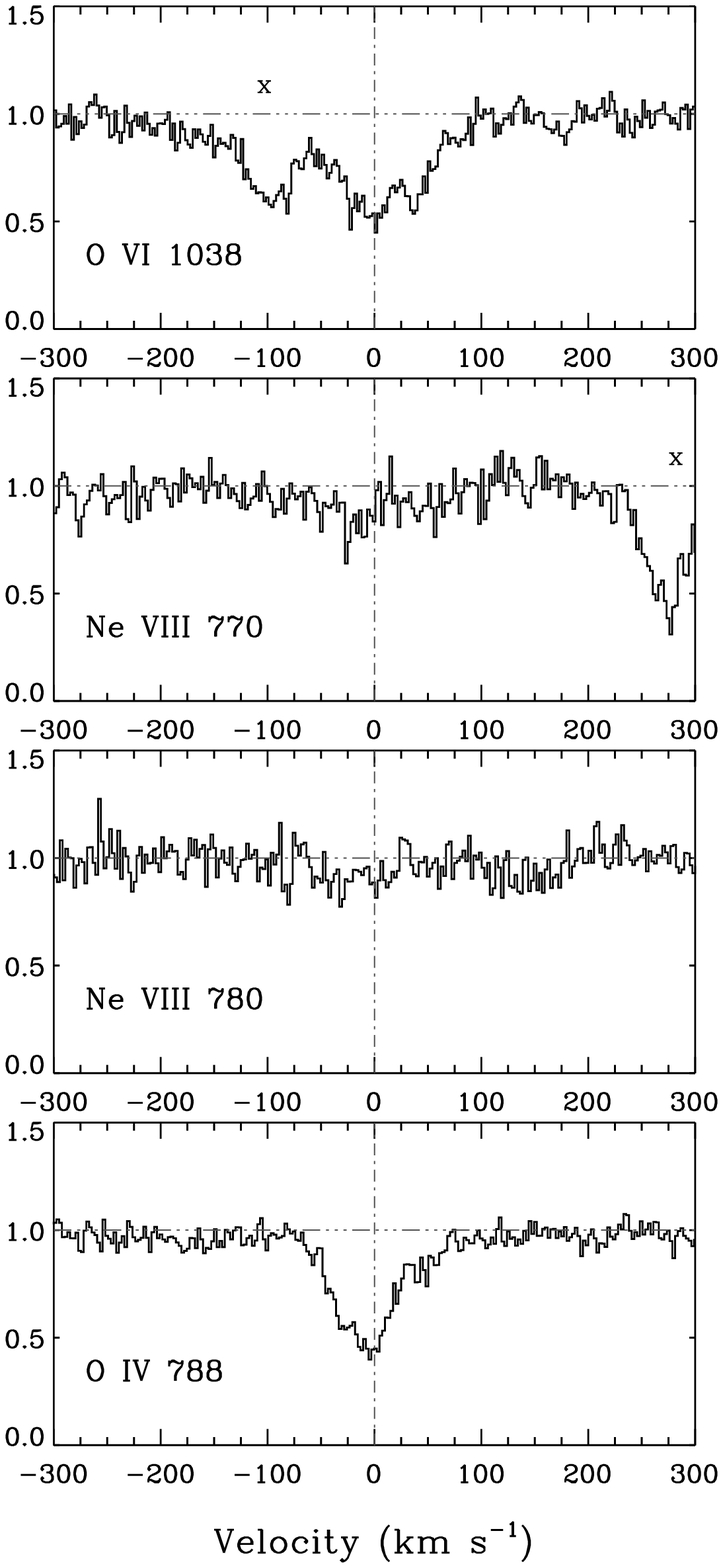}
\end{center}
\protect
{\large Fig.~1a~--~Continuum normalized spectrum of PKS~$0405-123$ showing absorption profiles and wavelength regions of some prominent lines in the $z = 0.495096$ metal line absorption system. The display is in the rest-frame of the absorber, with $v = 0$~{\kms} corresponding to $z = 0.495096$. All panels are showing $HST$/COS spectrum, except {\OV}~$\lambda 630$ and {\Lya} which are covered by $FUSE$ and FOS observations respectively. The individual exposures used to produce the coadded COS spectrum are listed in table 1. {\Lyb}, {\OVIdblt}, {\NeVIIIdblt}, {\OIV}~$\lambda 788$, {\CIII}~$\lambda 977$, {\OIII}~$\lambda 833$ are COS detections of $> 3~\sigma$ significance. Features that are not part of the absorption system are marked "x" in each panel. The line measurements are listed in table 2.}
\label{fig:1a}
\end{figure*}

\clearpage
\begin{figure*}
\begin{center}
\includegraphics[scale=0.67]{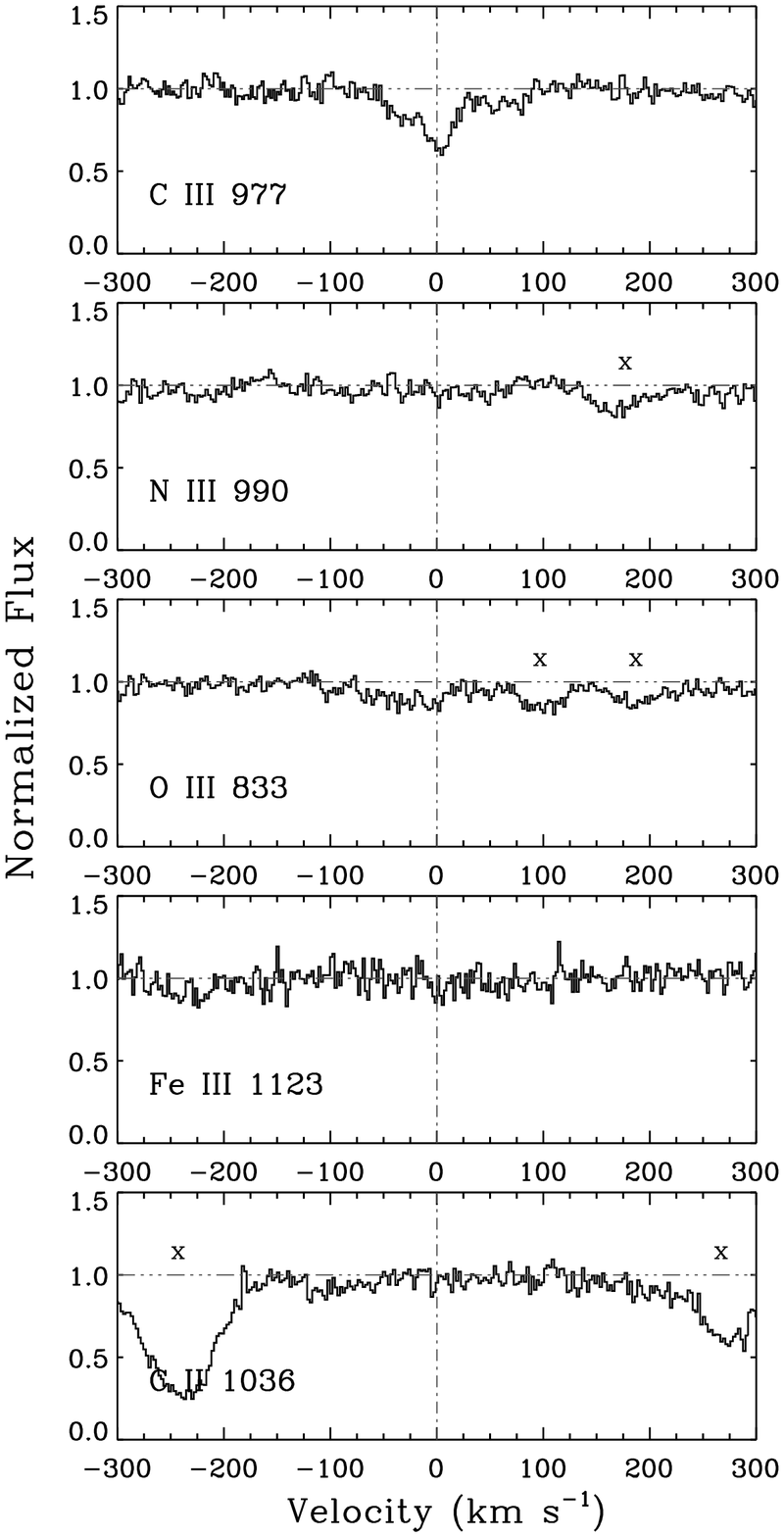}\hspace{0.1in}
\includegraphics[scale=0.67]{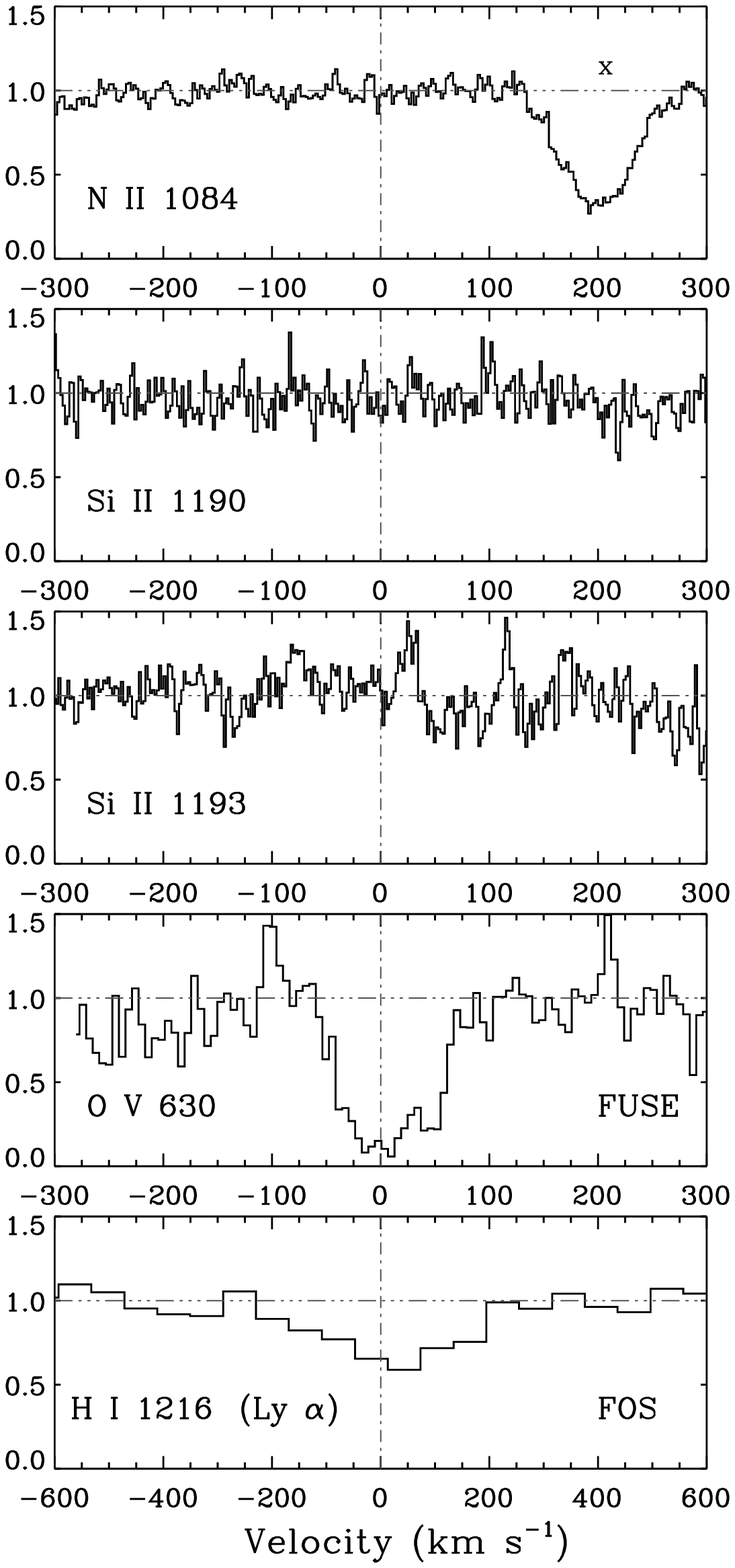}
\end{center}
\protect
{Fig.~1b~--~Continuation of Figure 1a.}
\label{fig:1b}
\end{figure*}

\setcounter{figure}{1}
\clearpage
%\begin{landscape}
%\begin{figure*}
\begin{sidewaysfigure}
\begin{center}
\includegraphics[scale=1.0,angle=90]{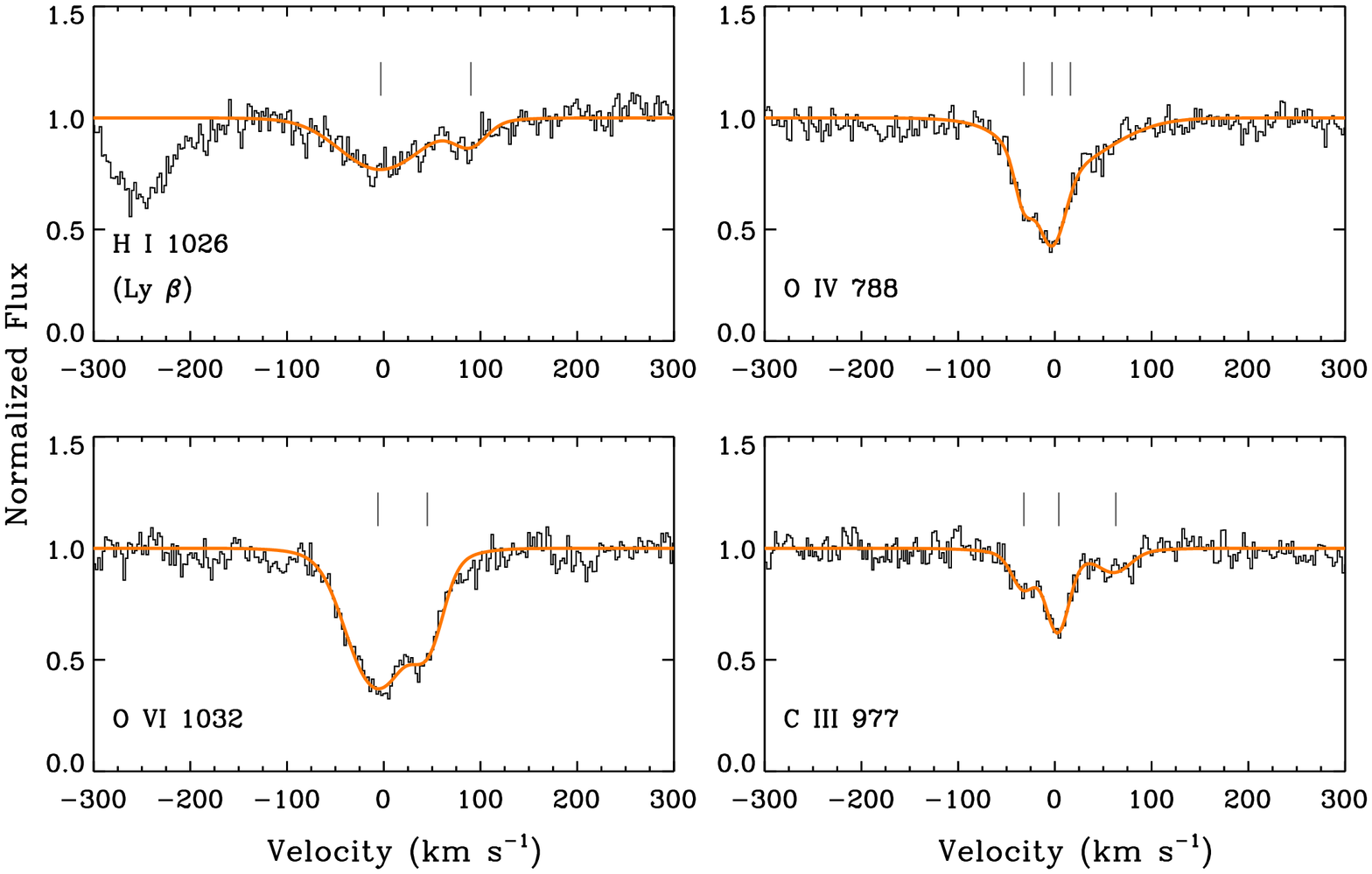}
\end{center}
\protect
\caption{\large Voigt profile fit models superimposed on the $z = 0.495096$ {\Lyb}, {\OVI}~$\lambda 1032$, {\OIV}~$\lambda 788$ and {\CIII}~$\lambda 977$ lines seen in the COS spectrum of PKS~$0405-123$. The profiles were fit using Fitzpatrick \& Spitzer (1997) routine. For each line, the model profiles were convolved with the COS instrumental spread function at the redshifted wavelength. The fit results are tabulated in Table 2. The centroid of the individual absorption components are marked by the vertical tick marks in each panel.}
\label{fig:2}
%\end{figure*}
%\end{landscape}
\end{sidewaysfigure}

\clearpage
\begin{figure*}
\begin{center}
\includegraphics[scale=0.58]{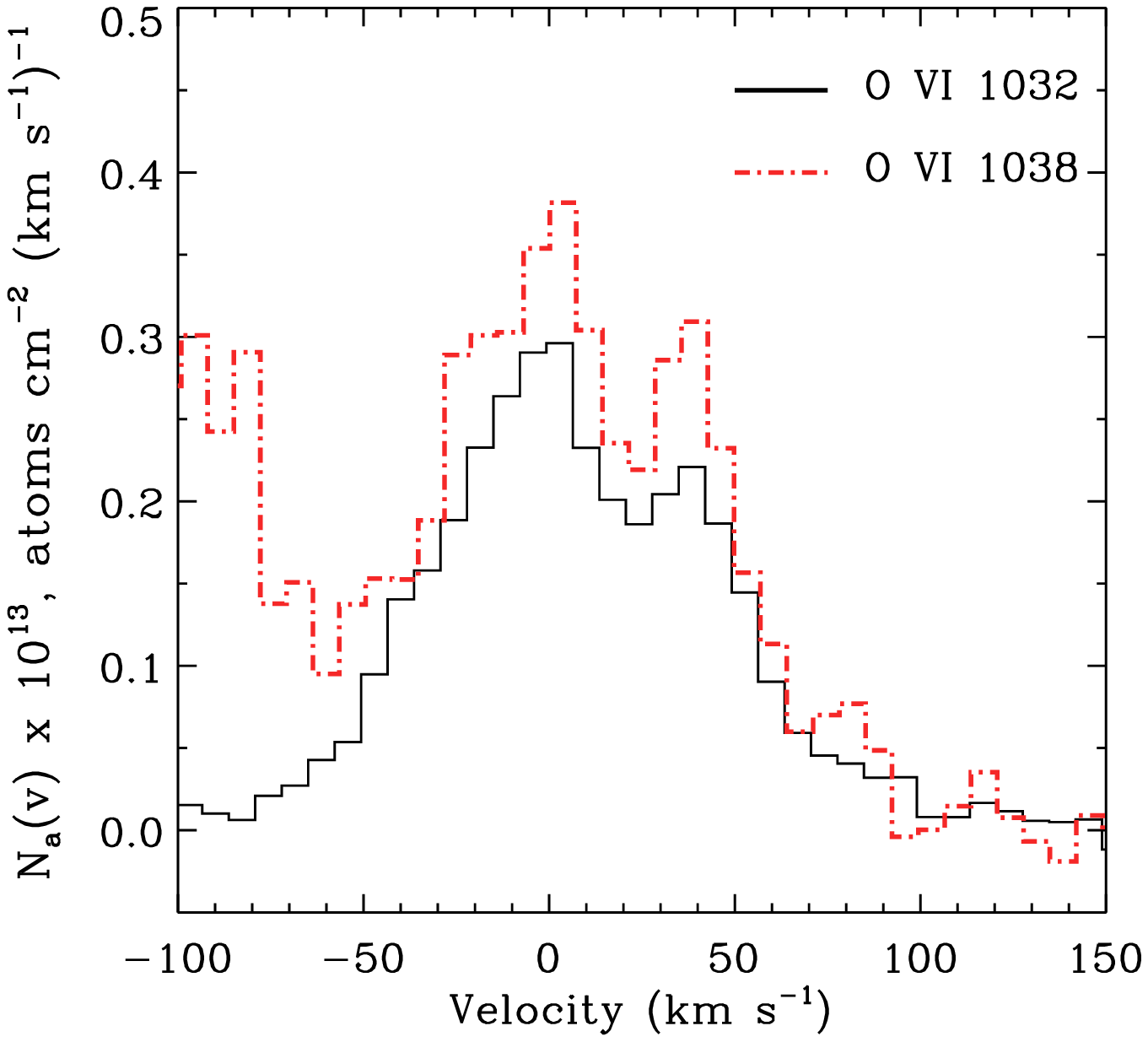}\vspace{0.15in}\hspace{0.2in}
\includegraphics[scale=0.58]{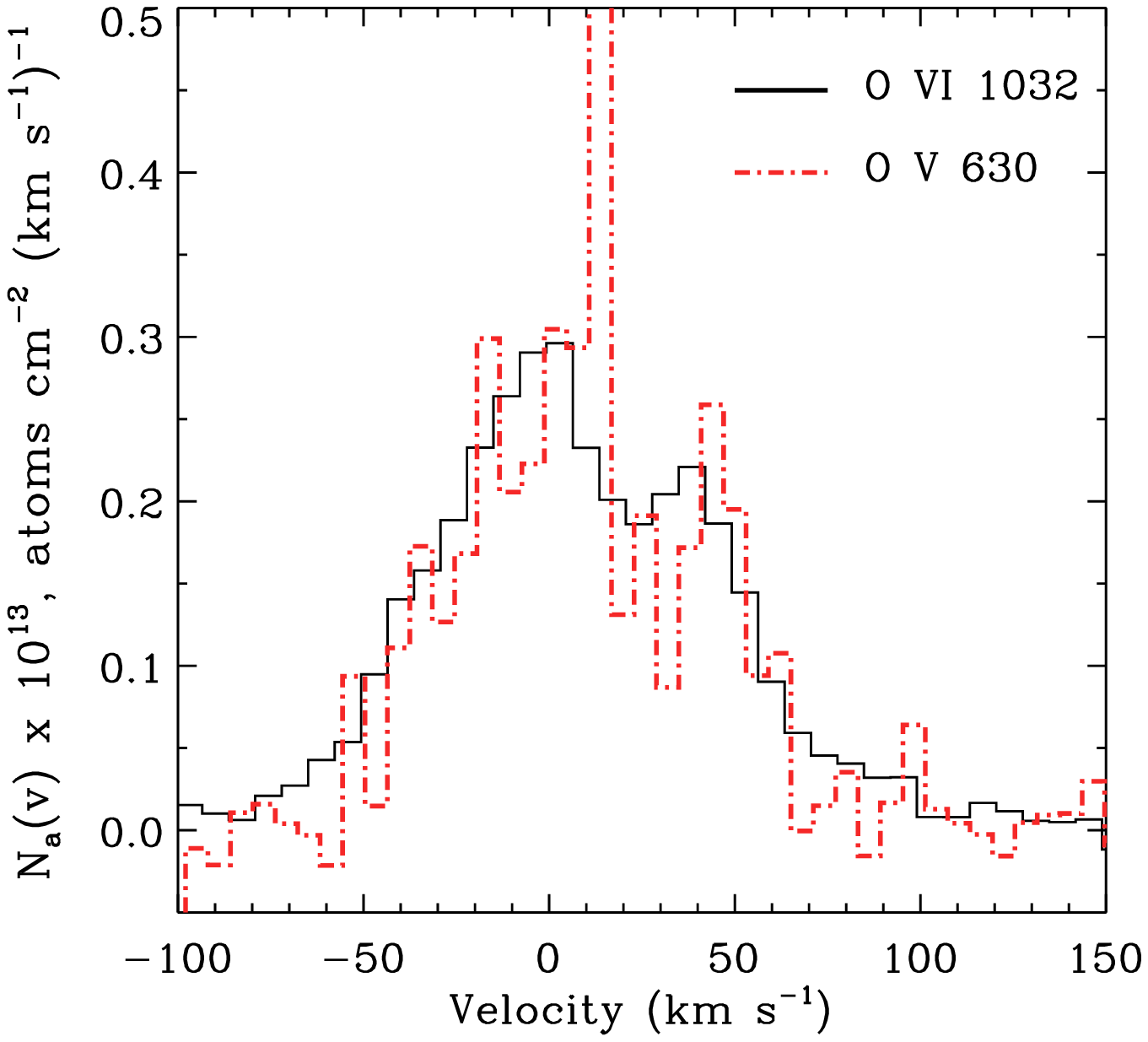}\vspace{0.15in}\hspace{0.2in}
\includegraphics[scale=0.58]{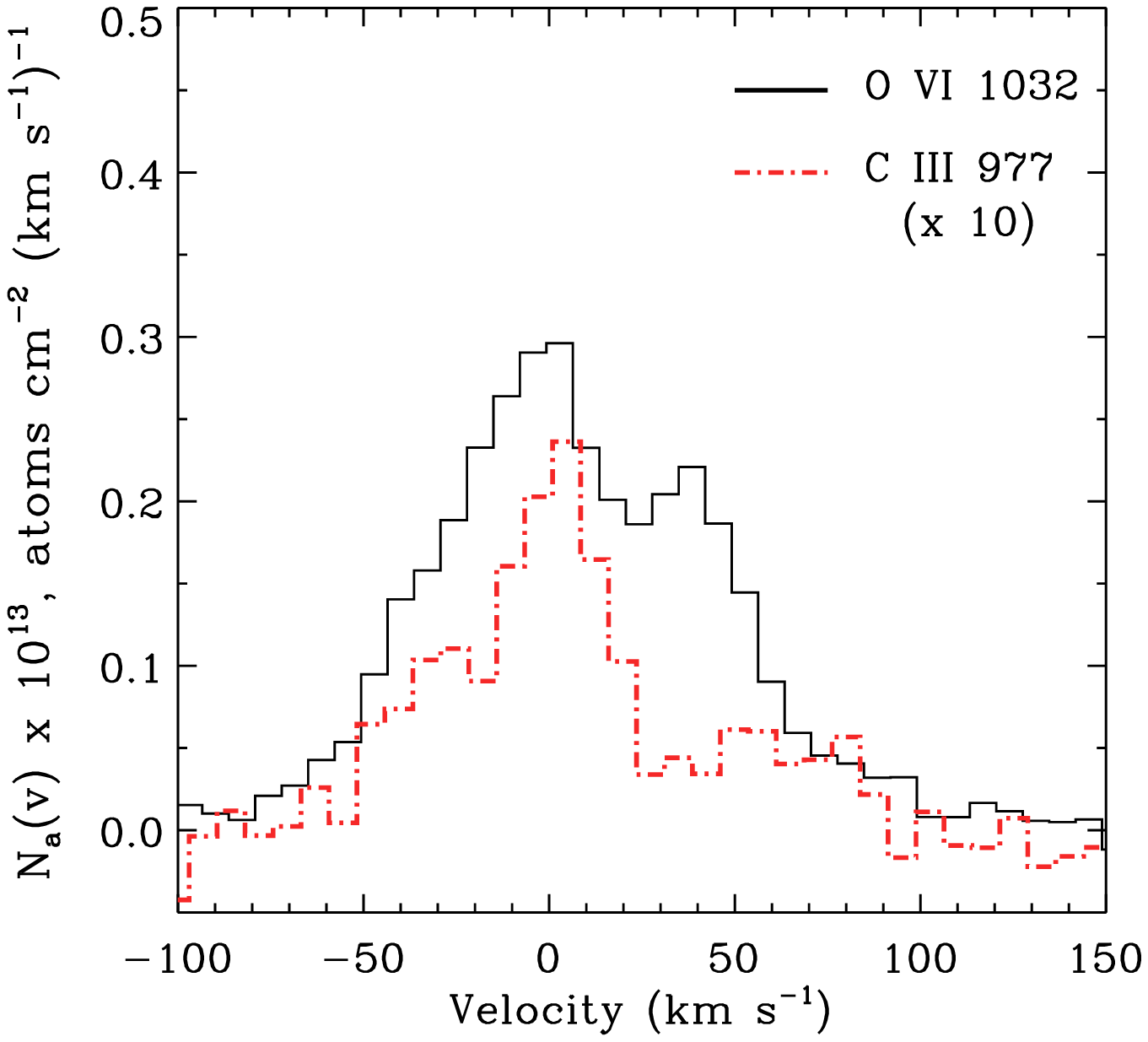}
\includegraphics[scale=0.58]{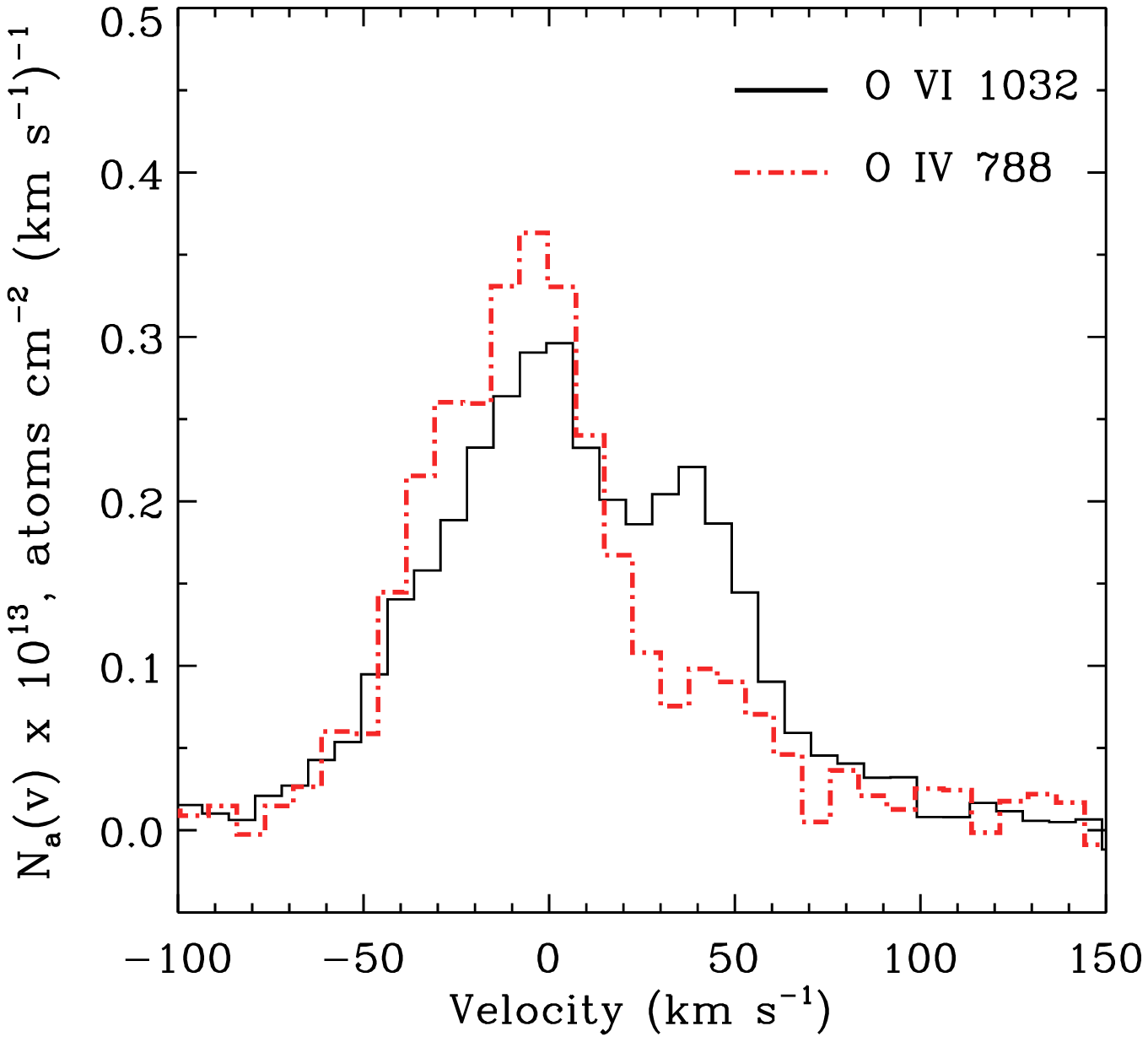}
\end{center}
\protect
\caption{\large Apparent column density $N_a(v)$ profiles of the {\OVIdblt} lines ({\it top left panel}), {\OV}~$\lambda 630$ compared with {\OVI}~$\lambda 1032$ ({\it top right panel}), {\OVI}~$\lambda 1032$ compared with {\CIII}~$\lambda 977$ ({\it bottom left panel}), {\OVI}~$\lambda 1032$ compared with {\OIV}~$\lambda 788$ ({\it bottom right panel}). The $N_a(v)$ profiles are displayed in the rest--frame of the $z = 0.495096$ absorber. All profiles except {\OV} are COS observations. The {\OV}~$\lambda 630$ is from $FUSE$. The difference in the apparent column densities between the individual lines of the {\OVI} doublet is indicative of unresolved saturation as discussed in section 4. The {\OVI}~$\lambda 1038$ is contaminated at $v < -50$~{\kms} by absorption unrelated to this system. The similarities between the component structures of {\OV} and {\OVI} is apparent and indicative of both ions predominantly tracing similar gas phase. The absorption in {\CIII} and {\OIV} in the velocity interval  $20 \leq v \leq 70$~{\kms} is weaker compared to {\OV} and {\OVI}. This could be due to higher ionization in the component of the gas producing the absorption.}
\label{fig:3}
\end{figure*}

\clearpage
%\begin{landscape}
%\begin{figure*}
\begin{sidewaysfigure}
\begin{center}
\includegraphics[scale=0.75,angle=90]{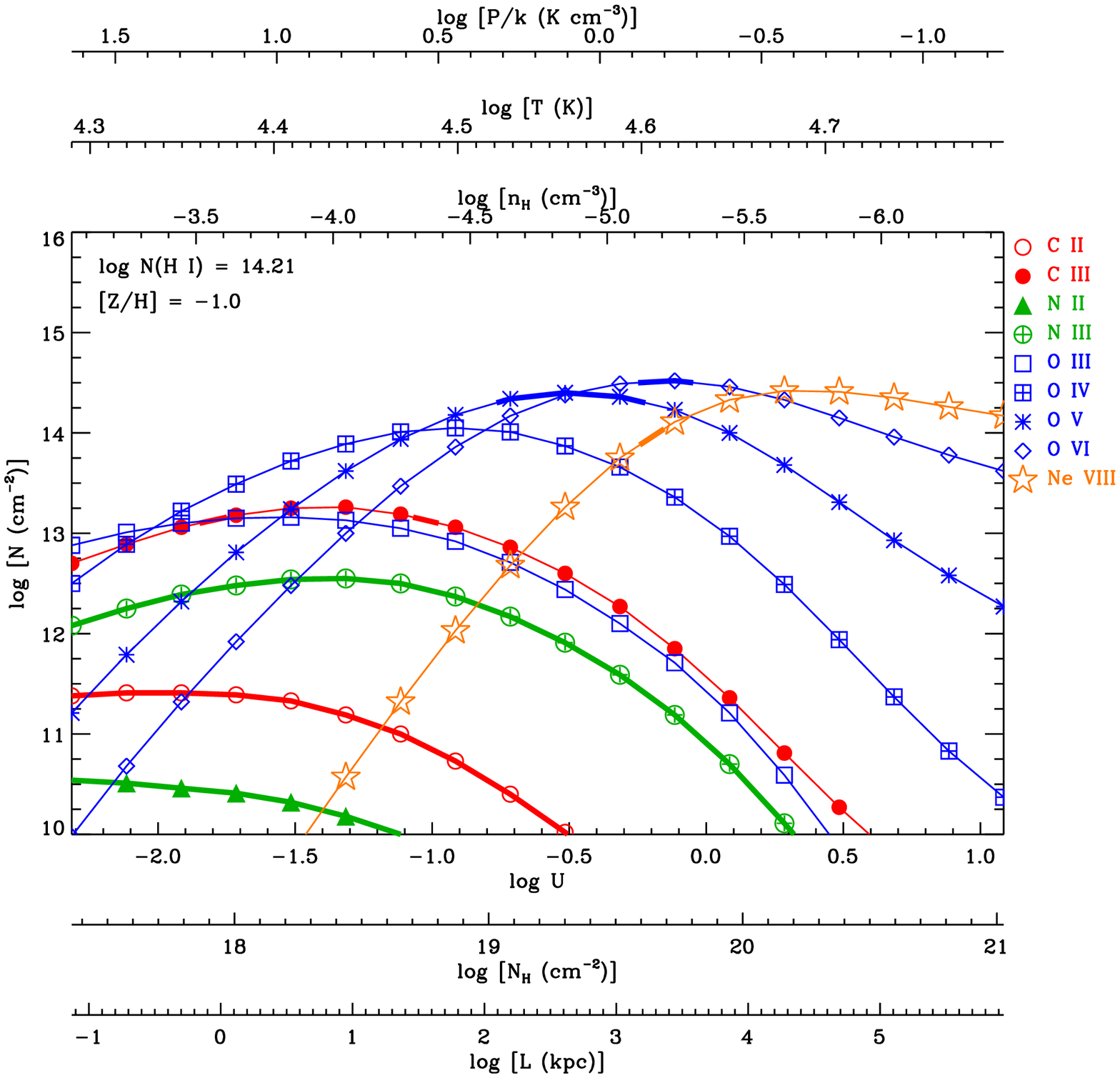}
\end{center}
\protect
\caption{Photoionization (PI) equilibrium models computed using Cloudy for log~$N(\HI) = 14.21$ and 10\% solar metallicity. We assume in these models the most recent solar relative elemental abundance estimations of \citet{asplund09}. The different curves show the column density predictions made by the PI models at different ionization parameter (log~$U$) values. The ionizing background is the Haardt \& Madau (2001) model for $z = 0.495096$ which includes UV photons from quasars and young star forming galaxies. The acceptable range of column densities for each ion, based on measurement, are highlighted in the photoionization curves using {\it thick} lines. The additional abscissa show the predicted values for the number density ($n_{\H}$), photoionization temperature (T),  pressure ($P/K$), total hydrogen column density ($N(\H)$), and path length ($L$). The column densities of {\OVI} and {\NeVIII} are simultaneously recovered for log~$U = -0.2$. The physical conditions predicted by this model are discussed in Sec 5.1.}
\label{fig:4}
%\end{figure*}
%\end{landscape}
\end{sidewaysfigure}

\clearpage
%\begin{landscape}
%\begin{figure*}
\begin{sidewaysfigure}
\begin{center}
\includegraphics[scale=0.8,angle=90]{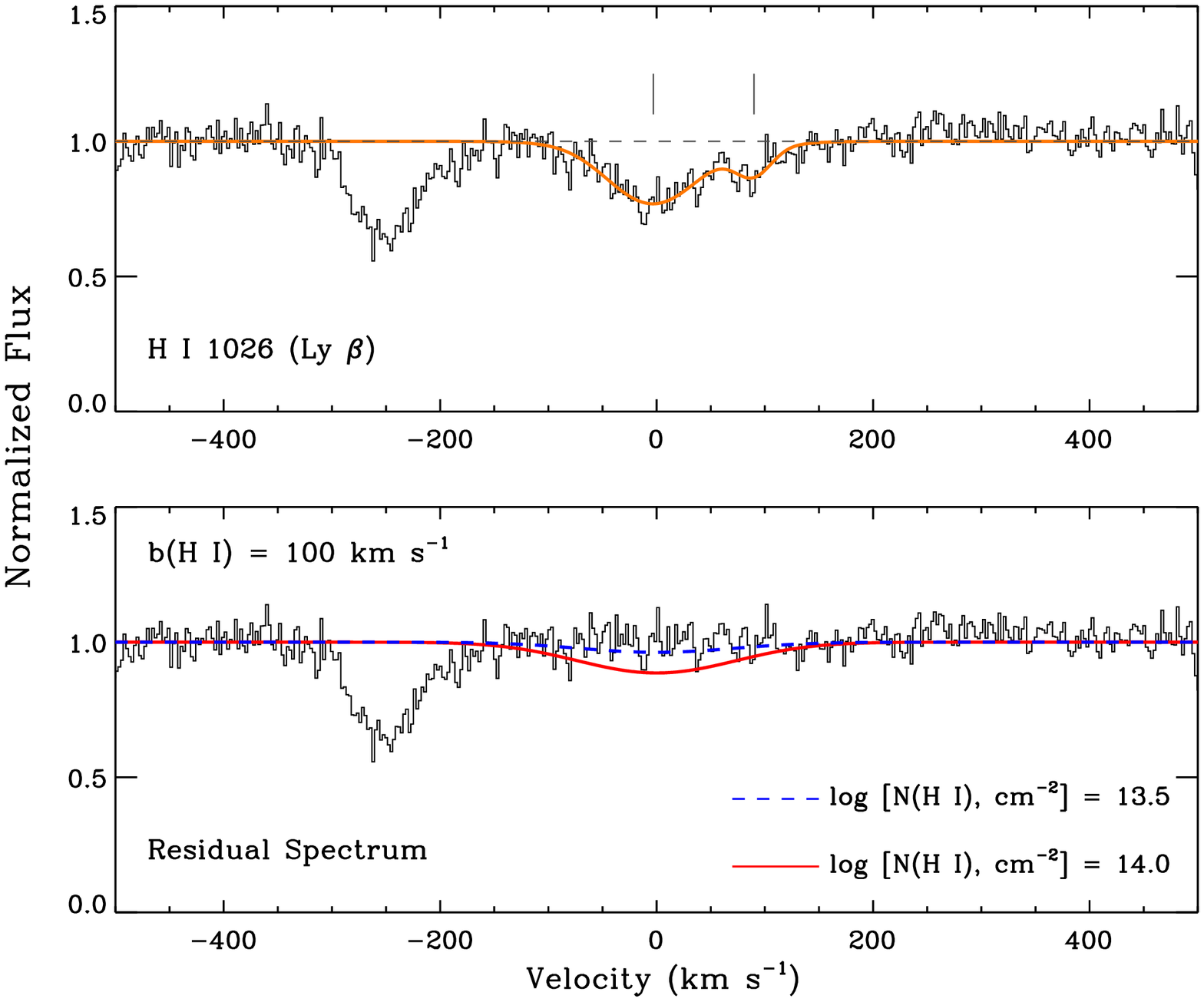}
\end{center}
\protect
\caption{\large The {\it top-panel} shows the $z = 0.495096$ rest-frame {\Lyb} absorption in the COS spectrum of PKS~$0405-123$. A Voigt profile fit to the {\HI} absorption is shown as the {\it thick line} is overlaid on the spectrum. The {\it bottom-panel} shows the residual absorption obtained after dividing the observed {\HI} absorption with the fit model. Superimposed on this residual spectrum are synthetic {\Lyb} profiles with $b = 100$~{\kms} for two different {\HI} column densities. The $b = 100$~{\kms} corresponds to the approximate thermal broadening to the {\HI} line corresponding to $T \geq 4.7 \times 10^5$~K, the temperature predicted by CIE models for the gas producing the {\NeVIII} absorption. Based on the figure in the {\it bottom- panel} we estimate the {\HI} column in the collisionally ionized gas to be log~$N(\HI) \lesssim 13.5$.}
\label{fig:5}
%\end{figure*}
%\end{landscape}
\end{sidewaysfigure}

\clearpage
\begin{figure*}
\begin{center}
\includegraphics[scale=0.75,angle=90]{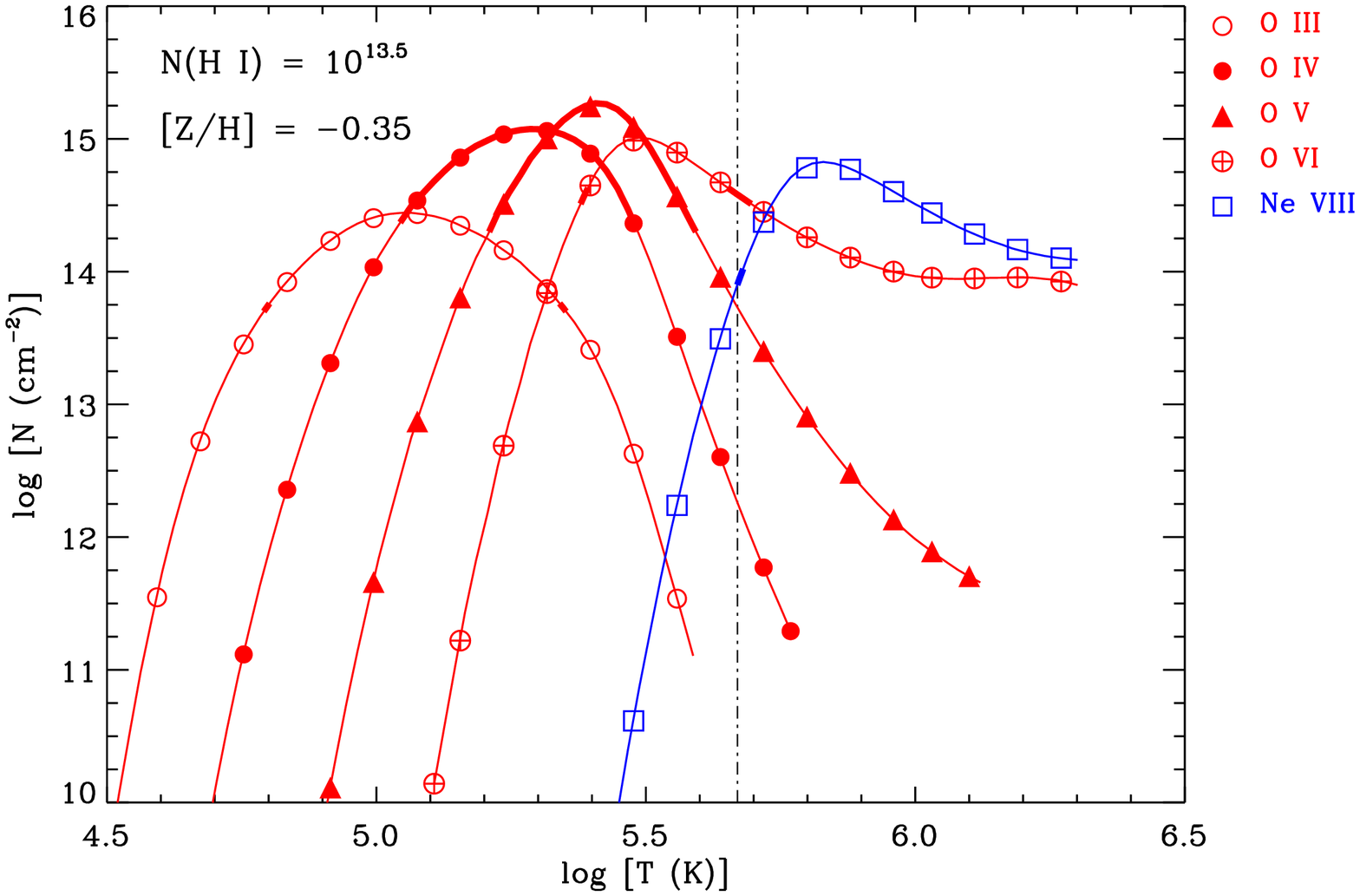}
\end{center}
\protect
\caption{\large Column densities predicted by the CIE models of \citet{gnat07} for {\NeVIII}, {\OVI}, {\OV}, {\OIV} and {\OIII} in the $z = 0.495096$ absorber, at the {\HI} column density of 13.5 dex and $-0.35$~dex metallicity. We assume in these models solar relative elemental abundances given by \citet{asplund09} which includes the most recent updates for Ne and O. The regions on the respective CIE curves where the measured column densities are recovered (within ${\pm}~1\sigma$ uncertainty) are highlighted using {\it thick} lines. The vertical {\it dotted} line marks $T = 4.7 \times 10^5$~K temperature at which the observed {\NeVIII} and {\OVI} column densities are recovered simultaneously by a single phase CIE model. This model can account for $\sim 30$\% of the observed {\OV} column density, but little {\OIV} or {\OIII} suggesting an origin for those ions in a separate gas phase.}
\label{fig:6}
\end{figure*}

\clearpage
\begin{figure*}
\begin{center}
\includegraphics[scale=0.9,angle=90]{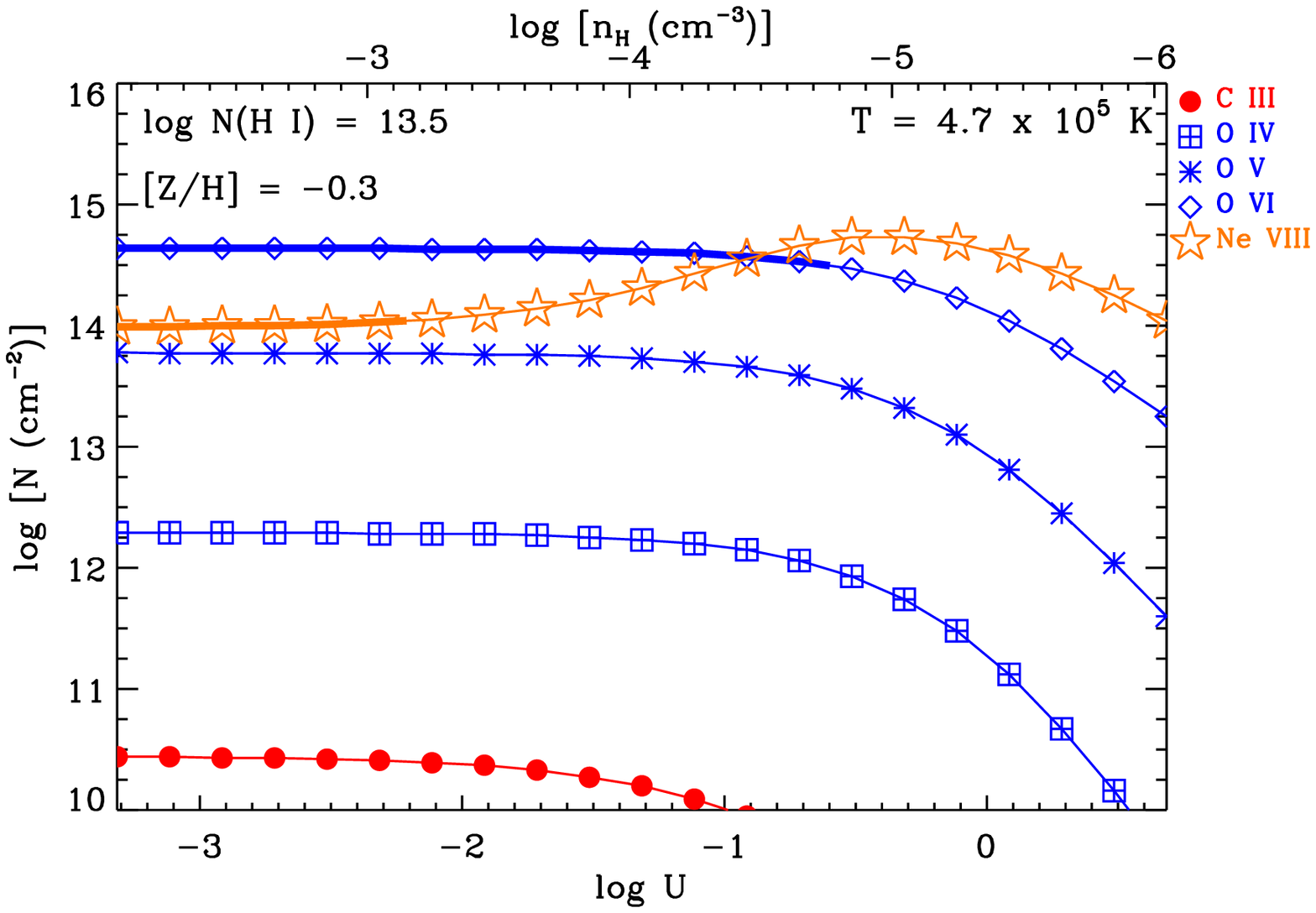}
\end{center}
\protect
\caption{\large Column density predictions from CIE and photoionization {\it hybrid} models computed using Cloudy for the {\NeVIII} phase of the absorber.  Solar relative elemental abundances given by \citet{asplund09} are assumed in these models. The temperature in the {\it hyrbid} models is fixed at  T $= 4.7 \times 10^5$~K  given by the $N(\NeVIII)/N(\OVI)$ at CIE (see Figure 6). The {\HI} column density is determined from the {\Lyb} profile (see Sec 5.2.1 and Figure 5). In the respective curves, the regions where the observed column densities are recovered (within ${\pm}~1\sigma$ uncertainty) are highlighted using {\it thick} lines. The column densities predicted for {\CII}, {\NII}, {\OIII} and {\NIII} are less than $10^{10}$~{\cmsq}, and hence are not displayed in the figure. The [Z/H] = $-0.3$ is the upper limit on metallicity. At that metallicity, for log~$U < -2.2$, {\NeVIII} and {\OVI} are simultaneously recovered from a single collisionally ionized phase. The physical conditions in the {\NeVIII} gas based on these {\it hybrid} models are discussed in Sec 5.2.2.}
\label{fig:6}
\end{figure*}


\begin{thebibliography}{XXX}

\bibitem[Asplund et al.(2009)]{asplund09} Asplund, M., Grevesse, N., Sauval, A.~J., \& Scott, P.\ 2009, \araa, 47, 481 

\bibitem[Bahcall et al.(1993)]{bahcall93} Bahcall, J.~N., Jannuzi, B.~T., Schneider, D.~P., \& Hartig, G.~F.\ 1993, \apj, 405, 491 

\bibitem[Bregman(2007)]{bregman07} Bregman, J.~N.\ 2007, \araa, 45, 221 

\bibitem[Brooks et al.(2009)]{brooks09} Brooks, A.~M., Governato, F., Quinn, T., Brook, C.~B., \& Wadsley, J.\ 2009, \apj, 694, 396 

\bibitem[Cen \& Ostriker(1999)]{cen99} Cen, R., \& Ostriker, J.~P.\ 1999, \apj, 514, 1 

\bibitem[Cen et al.(2001)]{cen01} Cen, R., Tripp, T.~M., Ostriker, J.~P., \& Jenkins, E.~B.\ 2001, \apjl, 559, L5 

\bibitem[Cen \& Fang(2006)]{cen06} Cen, R., \& Fang, T.\ 2006, \apj, 650, 573 


\bibitem[Cen \& Ostriker(2006)]{cen06} Cen, R., \& Ostriker, J.~P.\ 2006, \apj, 650, 560 


\bibitem[Chen \& Mulchaey(2009)]{chen09} Chen, H.-W., \& Mulchaey, J.~S.\ 2009, \apj, 701, 1219 

\bibitem[Bregman(2007)]{bregman07} Bregman, J.~N.\ 2007, \araa, 45, 221

\bibitem[Dahlen et al.(2005)]{dahlen05} Dahlen, T., Mobasher, B., Somerville, R.~S., Moustakas, L.~A., Dickinson, M., Ferguson, H.~C., \& Giavalisco, M.\ 2005, \apj, 631, 126 

\bibitem[Danforth \& Shull(2005)]{danforth05} Danforth, C.~W., \& Shull, J.~M.\ 2005, \apj, 624, 555 

\bibitem[Danforth et al.(2006)]{danforth06} Danforth, C.~W., Shull, J.~M., Rosenberg, J.~L., \& Stocke, J.~T.\ 2006, \apj, 640, 716 

\bibitem[Danforth \& Shull(2008)]{danforth08} Danforth, C.~W., \& Shull, J.~M.\ 2008, \apj, 679, 194

\bibitem[Danforth et al.(2010a)]{danforth10a} Danforth, C.~W., Stocke, J.~T., \& Shull, J.~M.\ 2010, \apj, 710, 613 

\bibitem[Danforth et al.(2010b)]{danforth10b} Danforth, C. W., Keeney, B. A., Stocke, J. T., Shull, J. M., \& Yao, Y 2010, \apj submitted, arXiv:1005:2191

\bibitem[Dav{\'e} et al.(2001)]{dave01} Dav{\'e}, R., et al.\ 2001, \apj, 552, 473 

\bibitem[Dixon et al.(2010)]{cos2010} Dixon, W. V., {\etal}2010, Cosmic Origins Spectrograph Instrument Handbook, Version 2.0 (Baltimore: STScI)

\bibitem[Ferland et al.(1998)]{ferland98} Ferland, G.~J., Korista, K.~T., Verner, D.~A., Ferguson, J.~W., Kingdon, J.~B., \& Verner, E.~M.\ 1998, \pasp, 110, 761

\bibitem[Fitzpatrick \& Spitzer(1997)]{fitzpatrick97} Fitzpatrick, E.~L., \& Spitzer, L., Jr.\ 1997, \apj, 475, 623 

\bibitem[Fox et al.(2004)]{fox04} Fox, A.~J., Savage, B.~D., Wakker, B.~P., Richter, P., Sembach, K.~R., \& Tripp, T.~M.\ 2004, \apj, 602, 738 

\bibitem[Fox et al.(2005)]{fox05} Fox, A.~J., Wakker, B.~P., Savage, B.~D., Tripp, T.~M., Sembach, K.~R., \& Bland-Hawthorn, J.\ 2005, \apj, 630, 332 

\bibitem[Froning \& Green(2009)]{froning09} Froning, C.~S., \& Green, J.~C.\ 2009, \apss, 320, 181 

\bibitem[Fukugita et al.(1998)]{fukugita98} Fukugita, M., Hogan, C.~J., \& Peebles, P.~J.~E.\ 1998, \apj, 503, 518 

\bibitem[Fukugita \& Peebles(2004)]{fukugita04} Fukugita, M., \& Peebles, P.~J.~E.\ 2004, \apj, 616, 643 

\bibitem[Ghavamian et al.(2009)]{coslsf} Ghavamian {\etal}2009, Preliminary Characterization of the Post- Launch Line Spread Function of COS, http://www.stsci.edu/hst/cos/documents/isrs/

\bibitem[Gnat \& Sternberg(2007)]{gnat07} Gnat, O., \& Sternberg, A.\ 2007, \apjs, 168, 213 

\bibitem[Green(2001)]{green01} Green, J.~C.\ 2001, \procspie, 4498, 229 

\bibitem[Haardt \& Madau(2001)]{haardt01} Haardt, F., \& Madau, P.\ 2001, Clusters of Galaxies and the High Redshift Universe Observed in X-rays.

\bibitem[Howk et al.(2009)]{howk09} Howk, J.~C., Ribaudo, J.~S., Lehner, N., Prochaska, J.~X., \& Chen, H.-W.\ 2009, \mnras, 396, 1875 

\bibitem[Jannuzi et al.(1998)]{jannuzi98} Jannuzi, B.~T., et al.\ 1998, \apjs, 118, 1 

\bibitem[Kere{\v s} et al.(2005)]{keres05} Kere{\v s}, D., Katz, N., Weinberg, D.~H., \& Dav{\'e}, R.\ 2005, \mnras, 363, 2 

\bibitem[Kere{\v s} \& Hernquist(2009)]{keres09} Kere{\v s}, D., \& Hernquist, L.\ 2009, \apjl, 700, L1

\bibitem[Kim et al.(1997)]{kim97} Kim, T.-S., Hu, E.~M., Cowie, L.~L., \& Songaila, A.\ 1997, \aj, 114, 1 

\bibitem[Lehner et al.(2006)]{lehner06} Lehner, N., Savage, B.~D., Wakker, B.~P., Sembach, K.~R., \& Tripp, T.~M.\ 2006, \apjs, 164, 1 

\bibitem[Lehner et al.(2007)]{lehner07} Lehner, N., Savage, B.~D., Richter, P., Sembach, K.~R., Tripp, T.~M., \& Wakker, B.~P.\ 2007, \apj, 658, 680 

\bibitem[Lehner et al.(2009)]{lehner09} Lehner, N., Prochaska, J.~X., Kobulnicky, H.~A., Cooksey, K.~L., Howk, J.~C., Williger, G.~M., \& Cales, S.~L.\ 2009, \apj, 694, 734 

\bibitem[Lockman \& Savage(1995)]{lockman95} Lockman, F.~J., \& Savage, B.~D.\ 1995, \apjs, 97, 1 

\bibitem[Mulchaey \& Chen(2009)]{mulchaey09} Mulchaey, J.~S., \& Chen, H.-W.\ 2009, \apjl, 698, L46 

\bibitem[Narayanan et al.(2009)]{narayanan09} Narayanan, A., Wakker, B.~P., \& Savage, B.~D.\ 2009, \apj, 703, 74 

\bibitem[Narayanan et al.(2010a)]{narayanan10a} Narayanan, A., Savage, B.~D., \& Wakker, B.~P.\ 2010, \apj, 712, 1443 

\bibitem[Narayanan et al.(2010b)]{narayanan10b} Narayanan, A., Wakker, B.~P., Savage, B.~D., Keeney, B.~A., Shull, J.~M., Stocke, J.~T.,  \& Sembach, K.~R.\ 2010, \apj accepted, arXiv:1008.2797 

\bibitem[Oppenheimer \& Dav{\'e}(2009)]{oppenheimer09} Oppenheimer, B.~D., \& Dav{\'e}, R.\ 2009, \mnras, 395, 1875 

\bibitem[Osterman et al.(2010)]{osterman10} Osterman, S., {\etal}2010, \apj, (in prep)

\bibitem[Penton et al.(2000)]{penton00} Penton, S.~V., Shull, J.~M., \& Stocke, J.~T.\ 2000, \apj, 544, 150 

\bibitem[Penton et al.(2004)]{penton04} Penton, S.~V., Stocke, J.~T., \& Shull, J.~M.\ 2004, \apjs, 152, 29 

\bibitem[Prochaska et al.(2004)]{prochaska04} Prochaska, J.~X., Chen, H.-W., Howk, J.~C., Weiner, B.~J., \& Mulchaey, J.\ 2004, \apj, 617, 718 

\bibitem[Prochaska \& Tumlinson(2009)]{prochaska09} Prochaska, J.~X., \& Tumlinson, J.\ 2009, Astrophysics in the Next Decade, Astrophysics and Space Science Proceedings, p.~419, 419, arXiv:0805.4635

\bibitem[Richter et al.(2004)]{richter04} Richter, P., Savage, B.~D., Tripp, T.~M., \& Sembach, K.~R.\ 2004, \apjs, 153, 165 

\bibitem[Richter et al.(2006)]{richter06a} Richter, P., Savage, B.~D., Sembach, K.~R., \& Tripp, T.~M.\ 2006, \aap, 445, 827 

\bibitem[Richter et al.(2006)]{richter06b} Richter, P., Fang, T., \& Bryan, G.~L.\ 2006, \aap, 451, 767 

\bibitem[Savage \& Sembach(1991)]{savage91} Savage, B.~D., \& Sembach, K.~R.\ 1991, \apj, 379, 245 

\bibitem[Savage et al.(2002)]{savage02} Savage, B.~D., Sembach, K.~R., Tripp, T.~M., \& Richter, P.\ 2002, \apj, 564, 631 

\bibitem[Savage et al.(2005)]{savage05} Savage, B.~D., Lehner, N., Wakker, B.~P., Sembach, K.~R., \& Tripp, T.~M.\ 2005, \apj, 626, 776 

\bibitem[Savage et al.(2010)]{savage10} Savage, B.~D., et al.\ 2010, \apj, 719, 1526 

\bibitem[Sembach \& Savage(1992)]{sembach92} Sembach, K.~R., \& Savage, B.~D.\ 1992, \apjs, 83, 147 

\bibitem[Sembach et al.(2004)]{sembach04} Sembach, K.~R., Tripp, T.~M., Savage, B.~D., \& Richter, P.\ 2004, \apjs, 155, 351 

\bibitem[Shull \& McKee(1979)]{shull79} Shull, J.~M., \& McKee, C.~F.\ 1979, \apj, 227, 131 

\bibitem[Shull(2009)]{shull09} Shull, J.~M.\ 2009, American Institute of Physics Conference Series, 1135, 301 

\bibitem[Stocke et al.(2006)]{stocke06} Stocke, J.~T., Penton, S.~V., Danforth, C.~W., Shull, J.~M., Tumlinson, J., \& McLin, K.~M.\ 2006, \apj, 641, 217 

\bibitem[Thom \& Chen(2008a)]{thom08a} Thom, C., \& Chen, H.-W.\ 2008, \apj, 683, 22 

\bibitem[Thom \& Chen(2008b)]{thom08b} Thom, C., \& Chen, H.-W.\ 2008, \apjs, 179, 37  

\bibitem[Tripp \& Savage(2000)]{tripp00} Tripp, T.~M., \& Savage, B.~D.\ 2000, \apj, 542, 42 

\bibitem[Tripp et al.(2000)]{tripp00} Tripp, T.~M., Savage, B.~D., \& Jenkins, E.~B.\ 2000, \apjl, 534, L1 

\bibitem[Tripp et al.(2006)]{tripp06} Tripp, T.~M., Aracil, B., Bowen, D.~V., \& Jenkins, E.~B.\ 2006, \apjl, 643, L77 

\bibitem[Tripp et al.(2008)]{tripp08} Tripp, T.~M., Sembach, K.~R., Bowen, D.~V., Savage, B.~D., Jenkins, E.~B., Lehner, N., \& Richter, P.\ 2008, \apjs, 177, 39 

\bibitem[Tumlinson \& Fang(2005)]{tumlinson05} Tumlinson, J., \& Fang, T.\ 2005, \apjl, 623, L97 

\bibitem[Tumlinson et al.(2005)]{tumlinson05} Tumlinson, J., Shull, J.~M., Giroux, M.~L., \& Stocke, J.~T.\ 2005, \apj, 620, 95 

\bibitem[Valageas et al.(2002)]{valageas02} Valageas, P., Schaeffer, R., \& Silk, J.\ 2002, \aap, 388, 741 

\bibitem[Verner et al.(1994)]{verner94} Verner, D.~A., Barthel, P.~D., \& Tytler, D.\ 1994, \aaps, 108, 287 

\bibitem[Wakker et al.(2003)]{wakker03} Wakker, B.~P., et al.\ 2003, \apjs, 146, 1

\bibitem[Wakker(2006)]{wakker06} Wakker, B.~P.\ 2006, \apjs, 163, 282

\bibitem[Wakker \& Savage(2009)]{wakker09} Wakker, B.~P., \& Savage, B.~D.\ 2009, \apjs, 182, 378 

\bibitem[Weymann et al.(1998)]{weymann98} Weymann, R.~J., et al.\ 1998, \apj, 506, 1 

\bibitem[Williger et al.(2006)]{williger06} Williger, G.~M., Heap, S.~R., Weymann, R.~J., Dav{\'e}, R., Ellingson, E., Carswell, R.~F., Tripp, T.~M., \& Jenkins, E.~B.\ 2006, \apj, 636, 631 

\bibitem[Wright(2006)]{wright06} Wright, E.~L.\ 2006, \pasp, 118, 1711 

\end{thebibliography}
\end{document}